\documentclass[10pt]{article} %
\usepackage[accepted]{tmlr}

\usepackage{amsmath,amsfonts,bm}

\def\1{\bm{1}}

\DeclareMathAlphabet{\mathsfit}{\encodingdefault}{\sfdefault}{m}{sl}
\SetMathAlphabet{\mathsfit}{bold}{\encodingdefault}{\sfdefault}{bx}{n}

\usepackage{hyperref}
\usepackage{url}
\usepackage{amsmath,amssymb,amsfonts}
\usepackage{amsthm}
\usepackage{mathtools}
\usepackage{pifont}
\usepackage{multirow}
\usepackage{subcaption}
\usepackage{bm}
\usepackage{booktabs}

\newcommand{\cmark}{\ding{51}} %
\newcommand{\xmark}{\ding{55}} %
\newcommand{\xmarkgray}{{\color{lightgray}\ding{55}}} %
\newtheorem{prop}{Proposition}
\newtheorem{lemma}{Lemma}
\newtheorem{thm}{Theorem}
\newtheorem{cor}{Corollary}
\newtheorem{definition}{Definition}
\newtheorem{assumption}{Assumption}
\newcommand{\pmark}{\rlap{$^*$}}
\title{Multiple Treatments Causal Effects Estimation with \\ Task Embeddings and Balanced Representation Learning}

\author{\name Yuki Murakami  \email yuuki.murakami.xg@nttdocomo.com \\
      \addr NTT DOCOMO, INC.
      \AND
      \name Takumi Hattori \email takumi.hattori.zw@nttdocomo.com \\
      \addr NTT DOCOMO, INC.
      \AND
      \name Kohsuke Kubota \thanks{Corresponding author.} \email kubota.kos.oy@yokohama-cu.ac.jp\\
      \addr Yokohama City University}

\def\month{11}  %
\def\year{2026} %
\def\openreview{\url{https://openreview.net/forum?id=xG2cjYSVjk}} %

\begin{document}

\maketitle
\begin{abstract}
The simultaneous application of multiple treatments is increasingly common in many fields, such as healthcare and marketing.
In such scenarios, it is important to estimate the single treatment effects and the interaction treatment effects that arise from treatment combinations.
Previous studies have proposed using independent outcome networks with subnetworks for interactions, or combining task embedding networks that capture treatment similarity with variational autoencoders.
However, these methods suffer from a lack of parameter sharing among related treatments, or from estimating unnecessary latent variables that reduce causal effect estimation accuracy.
To address these issues, we propose a novel deep learning framework that incorporates a task embedding network and a representation learning network with the balancing penalty.
The task embedding network enables parameter sharing across related treatment patterns by capturing treatment-effect-related similarities among treatment patterns.
The representation learning network with the balancing penalty learns representations nonparametrically from observed covariates while reducing discrepancies in representation distributions across different treatment patterns.
This process reduces estimation error associated with selection bias and reduces the risk of model misspecification.
Simulation studies demonstrate that the proposed method outperforms existing baselines, and application to real-world marketing datasets confirms the practical implications and utility of our framework.

\end{abstract}

\section{Introduction}
\label{sec:introduction}
Estimating the single and interaction effects of multiple, simultaneously applied treatments is a critical challenge in many fields, such as healthcare and marketing.
For example, combination drug therapies can lead to complex interactions and unforeseen side effects that are not apparent from studying each drug in isolation~\citep{gradman2010combination,webster2016combination,mokhtari2017combination}. 
Similarly, since the total impact of concurrent marketing promotions often exceeds the simple sum of their single effects, accounting for these interactions is crucial for optimizing marketing strategy~\citep{danaher2008effect,lemon2002developing,lesscher2021offline,naik2003understanding}. 
These situations demonstrate a critical need for robust analytical frameworks that can precisely estimate both single and interaction treatment effects to guide effective decision-making.

Whereas existing methods have been proposed to estimate causal effects under multiple treatments, they suffer from critical structural limitations that degrade their performance.
For example, Neural Counterfactual Relation Estimation~(NCoRE)~\citep{parbhoo2021ncore} addresses treatment interactions by using separate outcome prediction networks for each treatment and additional interaction subnetworks that are activated only when multiple treatments are applied simultaneously.
However, this reliance on separate networks prevents parameter sharing across similar treatments, which leads to unstable estimates, especially when samples for specific treatment patterns are limited. 
Similarly, Task Embedding–based Causal Effect Variational Autoencoder (TECE-VAE)~\citep{saini2019multiple}, which combines a treatment similarity–aware task embedding network with a variational autoencoder, assumes the existence of latent covariates and treats observed ones as proxies.
This approach, in turn, introduces a susceptibility to model misspecification because its forced estimation of latent variables can be detrimental when observed covariates are sufficient.
Therefore, the structural constraints of existing methods limit their ability to provide accurate and robust estimates of both single and interaction treatment effects.

In this study, we propose a novel deep learning framework called Causal Inference for Single and Interaction treatment effects Network~(CISI-Net).
CISI-Net integrates two core components: a task embedding network designed to capture treatment similarity and a representation learning network for reducing distributional discrepancies across treatment patterns.
First, the task embedding network learns to assign similar embedding vectors to similar treatment patterns, and it places treatment patterns with similar causal effects closer in the embedding space.
This alignment enables parameter sharing across related treatment patterns, which reduces the need for independent networks for each treatment pattern.
Moreover, it also improves the accuracy of causal effect estimation by leveraging treatment effect-related similarities among treatment patterns.
Second, the representation learning network with the balancing penalty learns representations nonparametrically from the observed covariates while aligning the representation distributions across different treatment patterns, which helps reduce estimation error associated with distributional imbalances caused by selection bias.
Unlike TECE-VAE, this data-driven flexibility removes the coercion to infer latent covariates, and thus reduces the risk of accuracy degradation caused by model misspecification.

To assess the effectiveness of the proposed framework, we conducted extensive evaluations using three simulations and two real-world datasets.
In three simulation datasets, our method outperformed all baseline approaches in estimating both single and interaction treatment effects.
Our real-world case study based on multiple marketing promotion datasets further demonstrates the practical implications and utility of our framework.

\section{Related Work}
This section reviews the main contributions and limitations of the deep learning–based methods for causal effect estimation.
These methods can broadly be categorized into two groups.
The first group consists of methods developed for single-treatment settings, which primarily aim to estimate the causal effect of one treatment applied in isolation. 
These methods can be adapted to multiple-treatment settings by treating each treatment pattern as a distinct treatment and are commonly used as baselines in multiple-treatment research~\citep{parbhoo2021ncore,saini2019multiple}.
The second group consists of methods developed for multiple-treatment settings.
These methods include approaches that extend single-treatment methods for multiple treatments or are specifically designed to handle multiple treatments.

Existing methods in the first group, designed for single-treatment settings~\citep{johansson2016learning,K_nzel_2019,  shi2019adapting, yang2024revisitingcounterfactualregressionlens, zhu2021direct}, are by their very nature unable to model the interaction effects that arise from the joint application of multiple treatments.
For example, Treatment-Agnostic Representation Network~(TARNet)~\citep{shalit2017estimating} learns a shared representation that is independent of treatment and uses it to predict counterfactual outcomes.
Counterfactual Regression~(CFR)~\citep{shalit2017estimating} extends TARNet by introducing a balancing penalty based on Integral Probability Metrics~(IPM)~\citep{sriperumbudur2010non}, which aligns the representation distributions of the treated and control groups to bound the expected estimation error.
Although these approaches are effective for estimating single treatment effects, they are fundamentally limited by their core assumption of treatment isolation, which makes them incapable of modeling interactions.

In the second group, existing methods for multiple treatments can be categorized according to their methodological paradigms.
These methods can be divided into data augmentation–based approaches~\citep{qian2021estimating} and model architecture–driven approaches~\citep{Chauhan_2025,parbhoo2021ncore,saini2019multiple, Tsuboi31122024,zou2020counterfactual}.
This distinction is important because data augmentation–based approaches rely on additional identification assumptions, such as sequential ignorability, which differ from the assumptions underlying model architecture-driven methods.
Even among model architecture–driven approaches, existing methods differ in their target settings, including the types of treatments, the number of outcomes, and the modeling assumptions.
Table~\ref{tab:related_work_comparison} summarizes these differences and clarifies the position of CISI-Net relative to other methods.
Although dosage combination network~(DC-Net)~\citep{schweisthal2023reliable} is architecturally similar to CISI-Net, it is specifically designed for continuous treatments and, therefore, cannot be directly applied to the binary treatment setting considered here.
Unlike H-Learner~\citep{Chauhan_2025}, which is designed to model treatment-outcome tasks across composite treatments and multiple outcomes, CISI-Net targets different estimands, which are causal effects under a single-outcome setting.
Accordingly, in what follows, we focus on previous studies that are most closely aligned with our setting of binary multiple-treatment effect estimation for a single outcome.

\begin{table}[t]
\centering
\caption{Comparison of model architecture-driven approaches under multiple treatments.}
\label{tab:related_work_comparison}
\renewcommand{\arraystretch}{1.02}
\resizebox{\textwidth}{!}{       
\begin{tabular}{lcccc}
\hline
Method & Treatment type & Outcome type & Parameter sharing mechanism & Latent covariate assumption \\ 
\hline
NCoRE 
& Binary 
& Single 
& None
& \xmark \\

TECE-VAE 
& Binary 
& Single 
& Task embedding + VAE 
& \cmark \\

H-Learner 
& Binary/Continuous 
& Multiple 
& Hypernetwork-based weight generation
& \xmark \\

DC-Net 
& Continuous 
& Single 
& Dosage-combination network 
& \xmark \\

CISI-Net 
& Binary 
& Single 
& Task embedding + shared outcome network
& \xmark \\
\hline
\end{tabular}
}
\end{table}

Within the class of model architecture-driven methods for single-outcome causal effect estimation, existing methods~\citep{parbhoo2021ncore,saini2019multiple,Tsuboi31122024, zou2020counterfactual} face two major limitations.
The first limitation is a structural inability to share parameters across related treatments, which leads to unstable estimation under data sparsity caused by the limited number of samples for specific treatment patterns.
The extensions of single-treatment methods can model interaction effects by assigning independent outcome prediction networks to each treatment pattern~\citep{parbhoo2021ncore,saini2019multiple}.
NCoRE introduces interaction subnetworks that are activated only when multiple treatments are applied simultaneously, which enables explicit modeling of interaction treatment effects.
These methods depend on having sufficient samples for each treatment pattern because each network is updated only with samples corresponding to its treatment pattern.
Since users who receive multiple treatments are typically rare~\citep{chu2022hierarchical}, networks corresponding to infrequent treatment patterns remain poorly trained.
In addition, because outcome prediction networks and interaction subnetworks are constructed independently, these methods cannot use similarities among treatments, which further increases instability in estimation.

The second limitation is a susceptibility to model misspecification.
Although TECE-VAE addresses the first limitation by combining a treatment similarity-aware task embedding network with a VAE, it relies on a VAE that assumes the presence of latent covariates.
This method assumes the presence of latent covariates, which are a subset of or the entirety of the true covariates that remain unobserved by the analyst, and treats observed covariates as proxies for these latent covariates.
When the assumed latent-variable generative model does not align with the true data-generating process, the inferred latent variables may fail to represent covariate structure, which leads to model misspecification.
A similar risk of model misspecification arises in other latent-variable–based approaches for multiple treatments~\citep{ma2021multi}, such as Variational Sample Re-weighting~\citep{zou2020counterfactual} and Deconfounder~\citep{wang2021proxy}.
Such assumptions about latent variables can reduce robustness and limit the practical applicability of these methods in real-world settings.

Our CISI-Net addresses these limitations through two core components.
First, the task embedding network captures treatment similarities by mapping related treatment patterns to proximal points in an embedding space.
By bringing treatment patterns with similar treatment effects closer together, it enables parameter sharing across treatment patterns, which improves estimation stability and removes the need for separate outcome networks for each treatment combination.
Second, the representation learning network with the balancing penalty directly learns balanced representations from observed covariates to reduce distributional discrepancies across treatment patterns, thereby controlling a source of estimation error.
Crucially, because this data-driven approach does not rely on the rigid latent-variable assumptions like TECE-VAE, this design minimizes the risk of accuracy degradation from model misspecification.

\section{Causal Inference under Multiple Treatments}
\label{sec:causal-inference}
We formulate the causal inference problem under multiple treatments within the potential outcomes framework~\citep{rubin2005causal}. 
The goal is to estimate the single and interaction treatment effects of multiple binary treatments on a continuous outcome using observed covariates.
We consider $N$ independent units, indexed by $i = 1, \dots, N$.
For each unit $i$, we observe a covariate vector $\boldsymbol{x}_i \in \mathbb{R}^d$ drawn from the covariate space $\boldsymbol{X}$.
There are $K$ binary treatments.
For each unit $i$, let the random variable $\boldsymbol{T}_i$ denote the treatment assignment.
Its realization, $\boldsymbol{t}_i \in \{0,1\}^K$, is the vector representing the specific treatment the unit actually received.
Under any possible treatment vector $\boldsymbol{t} \in \{0,1\}^K$, the potential outcome for unit $i$ is a scalar value $Y_i(\boldsymbol{t}) \in \mathbb{R}$, which represents the outcome that would have been observed had unit $i$ received treatment $\boldsymbol{t}$.
In practice, we only observe one of these potential outcomes for each unit.
The observed outcome is denoted by $Y_i$, and the specific realized outcome is denoted by $y_i$.
Therefore, for each unit $i$, the observed data consists of the triplet $(\boldsymbol{x}_i, \boldsymbol{t}_i, y_i)$.

To identify causal effects of interest from the observed data, we adopt the following three assumptions commonly used in observational studies~\citep{imbens2015causal}. 

\noindent\textbf{Assumption 1.~(Stable Unit Treatment Value Assumption)} \textit{
(1) no interference, meaning that the outcome of one unit is unaffected by the treatment assignments of other units; and (2) consistency of treatment, meaning that the potential outcomes correspond to well-defined and unique treatments~(i.e., $Y_i = Y_i(\boldsymbol{T}_i)$).}

\noindent\textbf{Assumption 2. ~(Ignorability.)} \textit{
For any treatment pattern, the potential outcome is independent of the assigned treatment $T$ given the observed covariates $X$.
Formally, for all $\boldsymbol{t}$, $$ Y(\boldsymbol{t}) \perp \boldsymbol{T} \mid \boldsymbol{X} $$
}

\noindent\textbf{Assumption 3. ~(Overlap.)} \textit{
Every unit has a non-zero probability of receiving any treatment pattern given its observed covariates.
Formally, for all $\boldsymbol{t}$ and $\boldsymbol{x}$, $$ 0 < P(\boldsymbol{T}=\boldsymbol{t} \mid \boldsymbol{X}=\boldsymbol{x}) < 1 $$
}

Our estimands of interest, the average single effect~(ASE) and the average interaction effect~(AIE), are defined using the conditional expected potential outcome $\mu(\boldsymbol{x}, \boldsymbol{t})$ given by
\begin{equation}\label{mu_function}
    \mu(\boldsymbol{x},\boldsymbol t) \coloneqq \mathbb{E} [Y(\boldsymbol t)\mid\boldsymbol X=\boldsymbol x\bigr].
\end{equation}
Under Assumptions 1-3, this quantity is identified from observed data. 
This key result is summarized as follows.
\begin{prop}
\label{prop:main_identifiable}
Under Assumptions 1-3, the conditional average potential outcome $\mu(\boldsymbol{x}, \boldsymbol{t})$ is identified and is equal to the conditional expectation of the observed outcome as follows:
\begin{equation}
\mu(\boldsymbol{x}, \boldsymbol{t}) = \mathbb{E}[Y \mid \boldsymbol{X} = \boldsymbol{x}, \boldsymbol{T} = \boldsymbol{t}].
\end{equation}
\end{prop}

We now formally define our estimands of interest.
First, the ASE for treatment $k$ is the average effect of applying only treatment $k$ compared to no treatment.
Let $\boldsymbol{t}_{+k}$ be the one-hot vector for treatment $k$.
The ASE is defined as
\begin{equation}\label{ASE}
\tau_{\mathrm{ASE}}(k) \coloneqq 
\mathbb{E}_{\boldsymbol{X}} \left[
\mu(\boldsymbol{x}, \boldsymbol{t}_{+k}) - 
\mu(\boldsymbol{x}, \boldsymbol{0}) \right].
\end{equation}
The ASE is mathematically equivalent to the average combination effect~\citep{egami2019causal} under the special case where the treatment vector is one-hot, meaning that only a single treatment is applied.
The ASE captures the causal effect of applying a single treatment.

Second, the AIE~\citep{egami2019causal} quantifies the interaction effect for a combination of treatments $S$~(where $S \in \{ S' \subseteq \{1,\dots,K\} \mid |S'| \ge 2 \}$) as follows:
\begin{equation}\label{AIE}
\tau_{\mathrm{AIE}}(S) \coloneqq 
\mathbb{E}_{\boldsymbol{X}} \left[
\sum_{Q \subseteq S} (-1)^{|S| - |Q|} \,
\mu\left(\boldsymbol{x}, \boldsymbol{t}_{(+Q)}\right) \right],
\end{equation}
where ${t}_{(+Q)}$ is the treatment vector that sets the components indexed by $Q$ to one and all remaining components to zero.
This definition corresponds to the special case where the baseline treatment vector is the all-zero condition and each treatment element is binary, as discussed in prior work~\citep{egami2019causal}.
The AIE defines interaction effects as differences that cannot be explained by simply adding single treatment effects.
This additive-scale definition captures interaction effects in terms of absolute differences, without imposing any assumption that the outcome-generating process is additive.
For example, when $K=2$ and $ S =\{1, 2\}$, $\tau_{\mathrm{AIE}}(\{1,2\}) =\mathbb{E}_{\boldsymbol{x}}[\mu(\boldsymbol{x}, (1,1)) -\mu(\boldsymbol{x}, (1,0)) -\mu(\boldsymbol{x}, (0,1))+\mu(\boldsymbol{x}, (0,0))]$.
$\tau_{\mathrm{AIE}}(\{1,2\})$ quantifies how much the treatment effect of $t = (1,1)$ differs from what would be expected if the treatment effects of $(1,0)$ and $(0,1)$ were simply additive. 

A direct consequence of Proposition 1 is that our main estimands are also identified. 
The proofs for Proposition 1 and Corollary 1 are provided in Appendix~\ref{appendix:proofs_of_identifiability}. 
\begin{cor}
Under Assumptions 1-3, the ASE and AIE are identified.
\end{cor}

\section{Causal Inference for Single and Interaction Treatment Effects Network}
\label{sec:proposed_method}
\subsection{Model Architecture}
\begin{figure}[tb]
  \centering
  \includegraphics[width=0.76\linewidth, keepaspectratio]{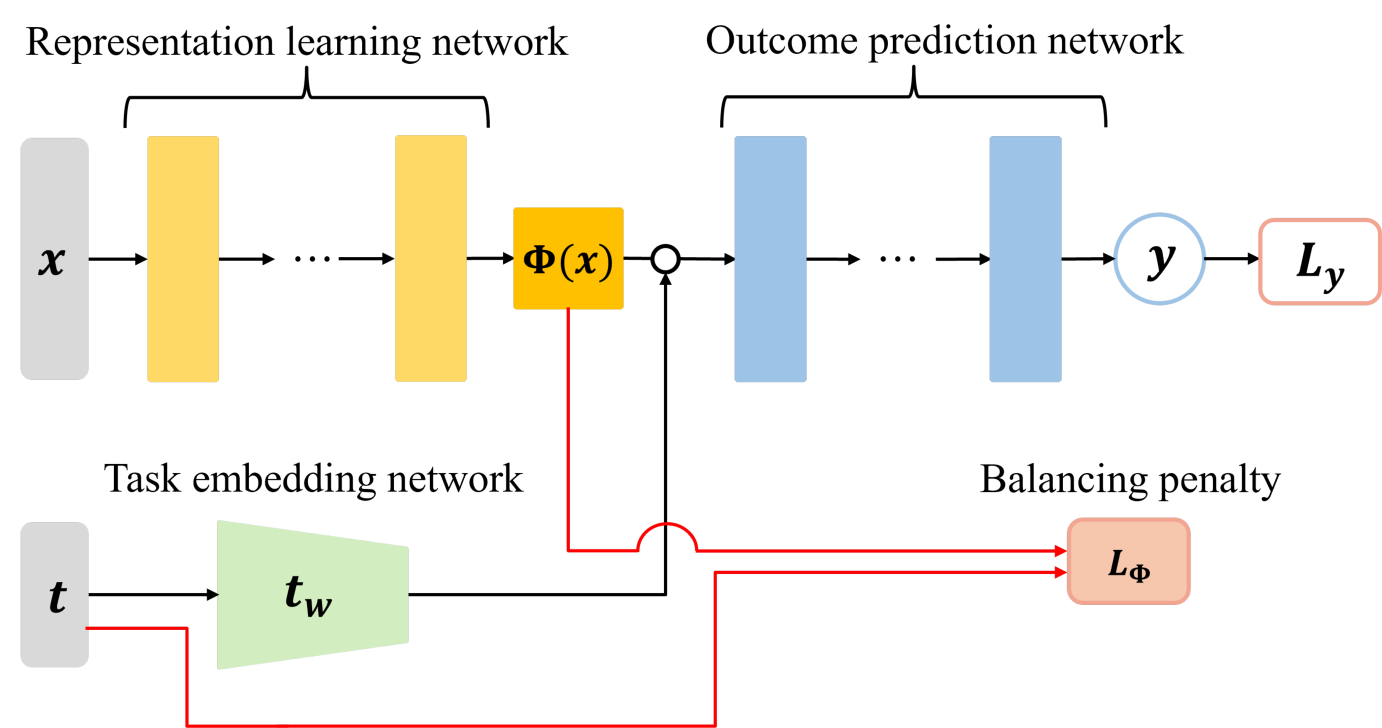}
  \caption{The architecture of CISI-Net consists of three components: the representation learning network~(yellow), the task embedding network~(green), and the outcome prediction network~(blue). The latent representation $\Phi(\boldsymbol{x})$ is concatenated with the task embedding vector $t_w(\boldsymbol{t})$ to predict the outcome $y$. The model is trained with two loss terms: the prediction loss $L_y$  and the balancing penalty $L_\Phi$~(red).}
  \label{fig:model_architecture}
\end{figure}

The objective of CISI-Net is to learn $\mu(\boldsymbol{x},\boldsymbol t)$ and estimate the causal effects defined in Equations \eqref{ASE} and \eqref{AIE} for any possible treatment vector $\boldsymbol t$. 
Figure~\ref{fig:model_architecture} illustrates the architecture of CISI-Net for estimating $\hat{\mu}(\boldsymbol{x}, \boldsymbol{t})$.
The model consists of three main components:~(i) a representation learning network with the balancing penalty, ~(ii) a task embedding network, and ~(iii) an outcome prediction network. 
These components are jointly optimized end-to-end, which ensures that all components are consistently aligned to learn $\mu(\boldsymbol{x},\boldsymbol{t})$ and reliably estimate the causal effects.

First, the representation learning network with the balancing penalty maps the observed covariates $\boldsymbol{x} \in \mathbb{R}^d$ to a latent representation space that is suitable for estimating causal effects with reduced distributional imbalances caused by selection bias.
Formally, it learns a function $\Phi: \mathbb{R}^d \rightarrow \mathbb{R}^{p}$, where $p$ is a hyperparameter indicating the dimensionality of the learned representations. 
A balancing penalty based on the IPM~\citep{shalit2017estimating} is applied to the learned representations to minimize the discrepancy between the distributions associated with different treatment patterns, which reduces distributional imbalances.
This helps reduce counterfactual prediction error arising from differences in covariate distributions across treatment patterns.

Second, the task embedding network enables the model to treat outcome prediction across treatment patterns as related prediction tasks by capturing similarities among treatment patterns, rather than modeling each treatment pattern independently.
Specifically, the task embedding network maps the binary treatment vector $\boldsymbol{t}$ into a $q$-dimensional continuous task embedding vector via a multi-layer perceptron $\mathrm{MLP}_w$~\citep{lecun2015deep}, denoted as $t_w(\boldsymbol{t}) = \mathrm{MLP}_w(\boldsymbol{t}) \in \mathbb{R}^q$.
This embedding captures similarities among treatments by positioning treatment patterns with related causal effects closer in the embedding space. 
As a result, the outcome prediction network, which receives the embedding vector $t_w(\boldsymbol{t})$, can exploit these similarities through parameter sharing across treatment patterns.

Third, the outcome prediction network takes the concatenated vector of $\Phi(\boldsymbol{x})$ and $t_w(\boldsymbol{t})$ as input, which forms a $(p+q)$-dimensional representation, and outputs the predicted outcome $Y$ using a neural network $h: \mathbb{R}^{p+q} \rightarrow \mathbb{R}$. 
Unlike previous approaches that use separate networks for each treatment~\citep{mondal2022memento,parbhoo2021ncore}, CISI-Net uses a single shared prediction function $h$ for all treatment patterns. 
In the task embedding space, because treatment patterns with similar causal effects are positioned close to each other, CISI-Net allows the outcome prediction network to reuse parameters across related treatment patterns.
Through this parameter sharing, CISI-Net achieves stable estimation by avoiding the construction of separate networks for each treatment pattern.

Our model estimates causal effects by predicting counterfactual outcomes based on the treatment vector as a direct input.
Through training, the proposed model learns a function $\hat{\mu}(\boldsymbol{x}, \boldsymbol{t})$ composed of a representation network $\Phi$, a task embedding network $t_w$, and an outcome prediction network $h$.
Instead of using the actual observed treatment vector $\boldsymbol{t}$, a counterfactual treatment vector $\boldsymbol{t}'$ is fed into $\hat{\mu}(\boldsymbol{x}, \boldsymbol{t})$, and the causal effect is estimated by Equations \eqref{ASE} and~\eqref{AIE}.

\subsection{Objective Function}

The objective function of CISI-Net provides a principled optimization strategy for simultaneously addressing two key challenges in causal effect estimation: maximizing outcome prediction accuracy and reducing distributional imbalances caused by selection bias.
Optimizing only for the former can result in biased counterfactual predictions, whereas focusing solely on the latter can compromise the expressive capacity of the model. 
To address this balance among accuracy, bias correction, and overfitting, we design the loss function $L$ as the sum of three components:
\begin{equation}
\label{eq:proposed_loss}
L = L_y + \alpha L_\Phi(\Phi, \boldsymbol{t}) + \beta \| w \|_2, 
\end{equation}
where the first term $L_y$ represents the outcome prediction error, the second term $L_\Phi(\cdot, \cdot)$ is a balancing penalty that reduces distributional differences in the representation space across treatment patterns, and the third term is an L2 regularization term applied to the network weights. 
The coefficients $\alpha$ and $\beta$ are hyperparameters that control the strength of the corresponding components.

Minimizing the outcome prediction error $L_y$ is essential for an accurate estimation of causal effects. 
When the frequency of observed treatment patterns is imbalanced, certain treatment patterns may be underrepresented, which results in biased predictions. 
To address this issue, we introduce a correction based on the empirical frequency of each treatment pattern~\citep{shalit2017estimating}:%
\begin{align}
    w_i(\boldsymbol{t}_i) &= \frac{1}{2} \left( \frac{1}{N} \sum_{j=1}^{N} \mathbb{I}[\boldsymbol{t}_i = \boldsymbol{t}_j] \right)^{-1}, \label{eq:weight} \\
    L_y &= \frac{1}{N} \sum_{i=1}^{N} w_i(\boldsymbol{t}_i) ( y_i - \hat{y}_i )^2, \label{eq:loss}
\end{align}
where $\hat{y}_i$ denotes the predicted outcome for unit $i$, and $w_i(\boldsymbol{t}_i)$ is the scaled inverse of the relative frequency of the treatment $\boldsymbol{t}_i$ in the observed data. 

The second term $L_\Phi$ is designed to reduce distributional discrepancies by aligning the representation distributions across treatment patterns.
Following the IPM-based representation-balancing framework~\citep{shalit2017estimating}, we extend representation balancing to the multi-treatment setting by computing the IPM for each unordered pair of treatment-pattern-specific representation distributions and averaging these discrepancies:
\begin{equation}
\label{eq:balancing_penalty_loss}
L_\Phi = \frac{1}{\binom{|T|}{2}} \sum_{\{\boldsymbol{a}, \boldsymbol{b}\} \in \binom{T}{2}} \mathrm{IPM} (\boldsymbol{R}_{\boldsymbol{a}}, \boldsymbol{R}_{\boldsymbol{b}}),
\end{equation}
where $\Phi(x)$ is the representation network mapping the input covariates $x$ into a latent representation space, and we define the set of representations for a given treatment pattern $\boldsymbol{a}$ as $\boldsymbol{R}_{\boldsymbol{a}} \coloneqq \{ \Phi(\boldsymbol{x}_i) \mid \boldsymbol{t}_i = \boldsymbol{a} \}$.
The $\mathrm{IPM}(\cdot,\cdot)$ measures the discrepancy between the two distributions of two such sets.
Here, $|T|$ denotes the number of distinct treatment patterns and $\binom{|T|}{2}$ is the total number of pairs of treatment patterns.

The theoretical role of this balancing penalty is established in Appendix~\ref{appendix:proof_upper_bounds}.
By extending the IPM-based representation-balancing framework from the two-group setting to multiple treatment patterns, we show that the pairwise balancing penalty controls the discrepancy term appearing in the upper bounds on the estimation errors of both ASE and AIE.
Because these bounds depend on both treatment-conditional factual prediction risks and the pairwise discrepancy term, this result provides a theoretical rationale for the objective in Equation~\eqref{eq:proposed_loss}: the prediction-loss term targets factual prediction accuracy, while $L_\Phi$ controls the distributional-discrepancy component relevant to ASE and AIE estimation.

\section{Simulation Experiment}
\label{sec:simulation_experiment}
This section describes the experimental setup, the hyperparameters for the proposed and baseline models, and the evaluation metrics used in the analysis.
To evaluate the effectiveness of the proposed method, we conducted two types of experiments. 
First, we compared the proposed method with several existing methods using three simulation datasets to assess the estimation performance. 
Second, we conducted an ablation study using one of the three simulation datasets to investigate the contributions of the key components of the proposed method, specifically the task embedding network and the balancing penalty.

\subsection{Simulation Datasets}
\label{subsec:simulation_datasets_gdp}
We design three simulation scenarios to comprehensively evaluate our framework's performance under different causal directed acyclic graphs and outcome-generating functions.
The first scenario, assuming full observability of true covariates, provides a fundamental setting for comparing estimation accuracy against baselines in the presence of treatment interactions.
The second scenario introduces latent covariates, which are true covariates but are unobservable.
This scenario is used as an empirical robustness analysis rather than as a setting directly covered by the identification assumptions in Section~\ref{sec:causal-inference}, because ignorability is not guaranteed in this setting.
Finally, the third scenario features multiple treatments but no interaction effects.
This setting is specifically designed to evaluate the robustness of CISI-Net against model misspecification by testing whether our model maintains performance even when the data lacks the interaction effects it is built to capture.
Collectively, these scenarios allow for a thorough assessment of our model's accuracy, empirical robustness to latent covariates, and resilience to model misspecification.

To clarify the words used in this section and beyond, we distinguish among three types of covariates: true covariates, latent covariates, and observed covariates.
First, the true covariates refer to the complete set of variables that influence both the treatment and the outcome. 
When all true covariates are observed, the data satisfy the assumption of no unobserved covariates.
Second, the latent covariates are a subset of or the entirety of the true covariates that are unobserved by the analyst.
They represent unobserved covariates that may induce bias in the estimation of causal effects if not properly accounted for~\citep{robins2000sensitivity}.
Third, the observed covariates consist of variables that are the true covariates themselves or proxy variables that partially capture the information of latent covariates.
The observed covariates represent the variables that are available to the analyst and are used as model inputs for estimation.

We vary the structure of the observed covariates $\boldsymbol{x}_{i,\text{o}}$ and the true covariates $\boldsymbol{x}_{i,\text{t}}$ by scenarios.
In all scenarios, only the observed covariates $\boldsymbol{x}_{i,\text{o}}$ are available to the model during training and evaluation. 
The true covariates $\boldsymbol{x}_{i,\text{t}}$ are used exclusively for the data-generating process~(i.e., to generate treatments and outcomes) and are treated as unobservable to the model at inference time.

In all scenarios, given the true covariates $\boldsymbol{x}_{i,\text{t}}$, treatment assignment and outcome generation follow the same functional form, with the number of treatments fixed at $K=3$ and the sample size set to $N=50,000$.
We introduce an indicator variable $H$ to mirror realistic conditions under which treatment assignment depends on true covariates in a non-smooth and discontinuous way~\citep{ascarza2018retention}.%
\begin{gather*}
H_i^{(1)} = \mathbb{I}\left(\,x_{i,\text{t}}^{(1)} + x_{i,\text{t}}^{(2)} > 1\,\right), \quad
H_i^{(2)} = \mathbb{I}\left(\,x_{i,\text{t}}^{(3)} + x_{i,\text{t}}^{(4)} < -0.5\,\right),\\
t_{i}^{(k)} \sim \mathrm{Bern} \left( \sigma \left( \boldsymbol{w}_{t_k}^\top \boldsymbol{x}_{i,\text{t}}-\lambda_1 H_i^{(1)} - \lambda_2 H_i^{(2)}- \delta \right) \right), \quad k=1,2,3,\\
\boldsymbol{t_i} = (t_{i}^{(1)},t_{i}^{(2)},t_{i}^{(3)}),\\
Y_i \sim \mathcal{N}\!\left(f\!\left(\boldsymbol{x}_{i,\mathrm{t}},\boldsymbol{t}_i,l\right),\,1^2\right),\quad l \in \{0,1\},
\end{gather*}
where $\mathbb{I}(\cdot)$ is the indicator function, $\sigma(\cdot)$ is the sigmoid function defined as $\sigma(x) = 1 / (1 + \exp(-x))$, and $\delta$ and $\lambda$ are bias parameters that vary across simulation scenarios.
The vector $\boldsymbol{w}_{t_k}$ is the weight vector whose elements are independently drawn from the uniform distribution $U(-0.5,0.5)$. 
The outcome-generating function $f$ is defined as follows:
\begin{align}
\begin{split}
\label{eq:dgp_outcome_generating_function}
f(\boldsymbol{x}_{i,\text{t}}, \boldsymbol{t_i}, l) 
&= \boldsymbol{w}_x^\top \boldsymbol{x}_{i,\text{t}}
 + (x_{i,\text{t}}^{(1)}+1)t_{i}^{(1)} 
 + 1.2(x_{i,\text{t}}^{(2)}+1)t_{i}^{(2)} \\
&\quad + 0.8(x_{i,\text{t}}^{(3)}+1) t_{i}^{(3)} 
+ l\Big\{ (x_{i,\text{t}}^{(4)}+0.5)t_{i}^{(1)} t_{i}^{(2)} \\
&\quad -0.5(x_{i,\text{t}}^{(5)}-0.1)t_{i}^{(1)} t_{i}^{(3)}
+0.1(x_{i,\text{t}}^{(6)}+0.1)t_{i}^{(2)} t_{i}^{(3)} \\
&\quad +0.7(x_{i,\text{t}}^{(7)}+1)t_{i}^{(1)} t_{i}^{(2)} t_{i}^{(3)}\Big\} + 2,
\end{split}
\end{align}
where vector $\boldsymbol{w}_{x}$ is the weight vector whose elements are independently drawn from the uniform distribution $U(-1,1)$, and the parameter $l \in \{0,1\}$ determines whether interaction effect terms are included ~($l=1$) or excluded ~($l=0$).

The first scenario assumes that true covariates are directly observable, as specified by the following equations: 
\begin{gather*}
x_{i,n}^{(j)} \sim N(c_n^{(j)}, 1^2), \quad 
x_{i,u}^{(j)} \sim U(-1,1), \quad \\
\boldsymbol{x}_{i,\text{o}} = \boldsymbol{x}_{i,\text{t}} = (\boldsymbol{x}_{i,n}, \boldsymbol{x}_{i,u}),
\end{gather*}
where $j \in \{1,\dots,15\}$ and $c_n^{(j)}$ are drawn from the uniform distribution $U(-1,1)$. In simulation dataset 1, $l=1$ , $\delta = 1$, $\lambda_1 = 1$ and $\lambda_2 = 1.2$.

The second scenario reflects the assumption adopted by deep generative models~\citep{louizos2017causal,saini2019multiple} that latent covariates exist.
Specifically, it considers real-world situations where latent covariates, such as the economic status or lifestyle of units, exist, and only proxy variables, which correspond to the observed covariates~(e.g., residential area, occupation, purchase history) are observed.
Ignorability conditional on the observed covariates is not guaranteed in this scenario.
Accordingly, Simulation 2 is not directly covered by the identification result under Assumptions~1--3 in Section~\ref{sec:causal-inference} and is interpreted only as an empirical robustness analysis.
Mathematically, this data-generating process is defined as:
\begin{gather*}
\boldsymbol{z}_i \sim N(\boldsymbol{0}, I_{10}), \quad
x_{i,n}^{(j)} \sim N(\boldsymbol{w}_n^{(j)\top} \boldsymbol{z}_i, 1^2), \\
x_{i,u}^{(j)} \sim N(\boldsymbol{w}_u^{(j)\top} \boldsymbol{z}_i, 5^2), \quad
x_{i,b}^{(j)} \sim \mathrm{Bern}(\sigma(\boldsymbol{w}_b^{(j)\top} \boldsymbol{z}_i)), \\
\boldsymbol{x}_{i,\text{o}} = (\boldsymbol{x}_{i,n}, \boldsymbol{x}_{i,u}, \boldsymbol{x}_{i,b}), \quad
\boldsymbol{x}_{i,\text{t}} = \boldsymbol{z}_i,
\end{gather*}
where $j \in \{1,\dots,10\}$ and the vectors $\boldsymbol{w}_n^{(j)}$, $\boldsymbol{w}_u^{(j)}$, and $\boldsymbol{w}_b^{(j)}$ are weight vectors whose elements are drawn independently from $U(-1,1)$. $I_d$ denotes the identity matrix $d \times d$, which is also used as the variance-covariance matrix. In simulation dataset 2, $l=1$ , $\delta = 0.8$, $\lambda_1 = 0.1$ and $\lambda_2 = 0.1$.

The third scenario adapts the structure of true and observed covariates to match that in the first scenario and assumes multiple treatments exist, with no interaction effects present.
In simulation dataset 3, the interaction treatment effect is eliminated by setting the interaction effect control parameter to $l=0$, and it is set to $\delta=1$, $\lambda_1 = 1$ and $\lambda_2 = 1.2$.

In summary, the three scenarios are designed to isolate different challenges.
Scenario 1 evaluates estimation accuracy under full observability.
Scenario 2 examines empirical robustness in a setting where the identification assumptions in Section~\ref{sec:causal-inference} are not guaranteed to hold.
Scenario 3 tests robustness to model misspecification in the absence of interaction effects.
Additional results for datasets generated under different settings~(e.g., homogeneous treatment effects and varying numbers of treatments) are provided in Appendix~\ref{appendix:additional_simulation_results}.

\subsection{Implementation Details}
\label{subsec:implement_details}

Our proposed method consists of three neural networks, all of which are built using fully connected~(FC) layers~\citep{lecun2015deep} with 200 units per hidden layer and leaky ReLU activation~\citep{xu2020reluplex}. 
All three neural networks have two hidden layers.
The task embedding network $t_w$ outputs a five-dimensional embedding vector. 
The balancing penalty coefficient $\alpha$ was set to 0.1, and the IPM used in the penalty was the Wasserstein distance~\citep{sriperumbudur2010non}.
During training, the Wasserstein distance is computed within each mini-batch using the treatment-pattern-specific empirical representation distributions.
Specifically, the learned representations are grouped according to the observed treatment patterns in each mini-batch, and the Wasserstein distance is evaluated for each unique unordered pair of treatment patterns only when both groups contain at least two samples.
The balancing penalty is then computed as the arithmetic mean of the Wasserstein distances over all eligible pairs and is set to zero when no eligible pair exists.

To enable a comprehensive comparison between methods that extend single treatment models to multiple treatment settings and those specifically designed for multiple treatments, we select five baseline methods: TARNet~\citep{shalit2017estimating}, CFR with Wasserstein-based balancing (CFR-WASS)~\citep{shalit2017estimating}, TECE-VAE~\citep{saini2019multiple}, NCoRE~\citep{parbhoo2021ncore}, and H-Learner~\citep{Chauhan_2025}. 
To adapt TARNet and CFR-WASS to the multi-treatment setting, we construct a separate outcome prediction network for each of the $2^K$ treatment patterns. 
In CFR-WASS, the balancing penalty coefficient $\alpha$ is fixed at one. 
In TECE-VAE, the latent dimension is set to 25, and the task embedding network has three hidden layers with 200 units and ELU activation~\citep{clevert2015fast}.
The task embedding network $t_w$ produces a five-dimensional embedding vector.
In NCoRE, each interaction subnetwork is implemented as two FC layers with 200 units per layer and ReLU activation~\citep{nair2010rectified}.
For H-Learner, we use a single-outcome variant in which the treatment vector and outcome index are separately mapped into embedding spaces and then concatenated to form a 64-dimensional task embedding.
The shared hypernetwork consists of two hidden layers with 100 ReLU units and generates the weights and biases of a task-specific target network with two hidden layers of 64 ReLU units.
Table~\ref{tab:parameter_count} shows the number of trainable parameters for each method as a reference for model capacity.

\begin{table}[t]
\centering
\caption{Comparison of the number of trainable parameters across methods.}
\label{tab:parameter_count}
\renewcommand{\arraystretch}{1.2}
\begin{tabular}{lcccccc}
\hline
 & TARNet & CFR-WASS & TECE-VAE & NCoRE & H-Learner & CISI-Net \\
\hline
\#Params & 0.73M & 0.73M & 0.43M & 0.28M & 0.64M & 0.14M \\
\hline
\end{tabular}
\end{table}

All models are trained with Adam optimizer~\citep{kingma2014adam}, using a learning rate of $10^{-4}$, a batch size of 256, and an L2 regularization of $10^{-5}$. 
Each dataset is divided into training sets of 70\% and test sets of 30\%, and all evaluations are carried out on the test set.
Training a model with one dataset and evaluating causal effects on an unseen test set is a conventional estimation scheme in causal inference using machine learning~\citep{liu2020estimating,okasa2022metalearnersestimationcausaleffects}.

\subsection{Evaluation Metrics}
We evaluate the estimation performance of single and interaction treatment effects using absolute errors. 
These are defined analogously to the absolute ATE estimation error commonly used in single-treatment studies~\citep{cheng2022evaluation,johansson2016learning}, and computed as the absolute difference between the true and estimated ASE or AIE. 
Specifically, we define $\epsilon_{\mathrm{ASE}}$ and $\epsilon_{\mathrm{AIE}}$ as follows:
\begin{equation}\label{ASEerror}
\epsilon_{\mathrm{ASE}}(k) = 
\left| 
\tau_{\mathrm{ASE}}(k) - \hat{\tau}_{\mathrm{ASE}}(k)
\right|,
\end{equation}
\begin{equation} \label{AIEerror}
\epsilon_{\mathrm{AIE}}(S)
= \left| \tau_{\mathrm{AIE}}(S) - \hat{\tau}_{\mathrm{AIE}}(S) \right|, 
\end{equation}
where $k\in \{1,\dots , K\}$ indexes single treatments, and $S$ denotes a subset of treatments with $|S| \ge 2$. $\hat{\tau}_{\mathrm{ASE}}(\cdot)$ and $\hat{\tau}_{\mathrm{AIE}}(\cdot) $ denote the estimated ASE and AIE.

To ensure robustness against randomness in data generation, we generate 100 independent datasets for each scenario using different random seeds. 
For both the baseline comparison and the ablation study, we evaluate each model across these 100 datasets and report the average of the resulting $\epsilon_{\mathrm{ASE}}$ and $\epsilon_{\mathrm{AIE}}$.

\section{Simulation Results}
\label{sec:simulation_results}
\paragraph{Overall Performance}
Table~\ref{tab:simulation_result_nonlinear_all} shows $\epsilon_{\mathrm{ASE}}$ and $\epsilon_{\mathrm{AIE}}$ of the proposed and baseline methods across three types of simulation datasets.
In all datasets, CISI-Net achieves the lowest or comparable estimation errors for most $\epsilon_{\mathrm{ASE}}$ and $\epsilon_{\mathrm{AIE}}$, emphasizing its superior ability to accurately estimate both single and interaction treatment effects under diverse conditions.
Notably, this performance was achieved despite CISI-Net having fewer trainable parameters than the baseline methods, as shown in Table~\ref{tab:parameter_count}.
In simulation scenario 2, where the observed covariates are only proxy variables for latent covariates, CISI-Net maintained high estimation performance.
These results demonstrate the empirical robustness of CISI-Net under the specific data-generating process where only proxy variables, rather than true covariates, are observed.
Furthermore, in simulation scenario 3, where no actual interaction treatment effects exist and the utility of the task embedding network is limited, CISI-Net still achieved the best accuracy.
These results suggest that CISI-Net is flexible, because it captures interaction treatment effects when they exist and avoids degradation when they do not.
This flexibility offers a significant practical advantage, because it frees analysts from the need to pre-screen for interaction treatment effects before applying CISI-Net.
In this way, CISI-Net shows stable performance across various data generation processes, whether interaction is present or absent, providing both accuracy and reliability compared to baselines.

\begin{table*}[tb]
\caption{Comparison of mean and standard deviation of $\epsilon_{\mathrm{ASE}}$ and $\epsilon_{\mathrm{AIE}}$ across simulation datasets. 
Here, $k \in \{1,2,3\}$ indexes single treatments for ASE, and $S \subseteq \{1,2,3\}$ (with $|S| \ge 2$) denotes treatment combinations for AIE.
An asterisk indicates that the estimation error is significantly different from that of CISI-Net based on a paired-sample $t$-test ($p<0.05$).
}
\label{tab:simulation_result_nonlinear_all}
\centering
\setlength{\tabcolsep}{6pt}          
\renewcommand{\arraystretch}{1.4}    
\resizebox{\textwidth}{!}{           
\begin{tabular}{p{0.4cm}lccc|cccc}
\hline
\multicolumn{2}{l}{} & \multicolumn{3}{c|}{$\epsilon_{\mathrm{ASE}}$} & \multicolumn{4}{c}{$\epsilon_{\mathrm{AIE}}$} \\
\cline{3-5}\cline{6-9}
Sim.&Method& $k=1$ & $k=2$ & $k=3$ & $S=\{1,2\}$ & $S=\{2,3\}$ & $S=\{1,3\}$ & $S=\{1,2,3\}$ \\
\hline
\multirow{6}{*}{\centering 1}
& TARNet 
& 0.12 $\pm$ 0.10 
& 0.12 $\pm$ 0.07
& 0.13 $\pm$ 0.07\pmark
& 0.18 $\pm$ 0.15\pmark  
& 0.16 $\pm$ 0.12 
& 0.18 $\pm$ 0.14\pmark 
& 0.34 $\pm$ 0.29\pmark \\

& CFR-WASS 
& 0.13 $\pm$ 0.09  
& 0.13 $\pm$ 0.08  
& 0.14 $\pm$ 0.06\pmark
& 0.19 $\pm$ 0.12\pmark  
& 0.17 $\pm$ 0.12  
& 0.17 $\pm$ 0.13\pmark  
& 0.33 $\pm$ 0.19\pmark \\

& NCoRE 
& \fontseries{b}\selectfont{0.10 $\pm$ 0.08} 
& 0.10 $\pm$ 0.08 
& 0.10 $\pm$ 0.08 
& 0.17 $\pm$ 0.12\pmark
& 0.16 $\pm$ 0.10  
& 0.17 $\pm$ 0.11\pmark  
& 0.23 $\pm$ 0.18\pmark \\

& TECE-VAE 
& 0.13 $\pm$ 0.10 
& 0.13 $\pm$ 0.10  
& 0.12 $\pm$ 0.09  
& \fontseries{b}\selectfont{0.12 $\pm$ 0.08} 
& 0.14 $\pm$ 0.10 
& 0.15 $\pm$ 0.11 
& 0.21 $\pm$ 0.15 \\

&  H-Learner
&  \fontseries{b}\selectfont{0.10 $\pm$ 0.07}
&  \fontseries{b}\selectfont{0.09 $\pm$ 0.07}
&  \fontseries{b}\selectfont{0.08 $\pm$ 0.06}
&  0.17 $\pm$ 0.13\pmark
&  0.14 $\pm$ 0.10
&  0.13 $\pm$ 0.10
&  0.23 $\pm$ 0.15\pmark \\

& \textbf{CISI-Net} 
& \fontseries{b}\selectfont{0.10 $\pm$ 0.07}
& \fontseries{b}\selectfont{0.09 $\pm$ 0.08} 
& \fontseries{b}\selectfont{0.08 $\pm$ 0.07} 
& \fontseries{b}\selectfont{0.12 $\pm$ 0.10} 
& \fontseries{b}\selectfont{0.13 $\pm$ 0.10} 
& \fontseries{b}\selectfont{0.12 $\pm$ 0.10} 
& \fontseries{b}\selectfont{0.12 $\pm$ 0.10} \\

\hline
\multirow{6}{*}{\centering 2}
& TARNet       
& 0.19  $\pm$ 0.14
& 0.27 $\pm$ 0.16\pmark    
& 0.14 $\pm$ 0.12
& 0.27 $\pm$ 0.16
& 0.23 $\pm$ 0.15
& 0.20 $\pm$ 0.14  
& 0.32 $\pm$ 0.26\pmark \\

& CFR-WASS     
& 0.20  $\pm$ 0.14
& 0.26 $\pm$ 0.16\pmark   
& 0.16 $\pm$ 0.13  
& 0.28 $\pm$ 0.17 
& 0.23 $\pm$ 0.17
& 0.22 $\pm$ 0.16  
& 0.36 $\pm$ 0.28 \\
 
& NCoRE        
& 0.32 $\pm$ 0.27\pmark  
& 0.35 $\pm$ 0.28\pmark  
& 0.30 $\pm$ 0.23\pmark  
& 0.31 $\pm$ 0.27  
& 0.36 $\pm$ 0.29\pmark  
& 0.36 $\pm$ 0.28\pmark  
& 0.55 $\pm$ 0.36\pmark \\

& TECE-VAE     
& 0.21  $\pm$ 0.17\pmark  
& 0.24 $\pm$ 0.19\pmark  
& 0.17 $\pm$ 0.13  
& 0.29 $\pm$ 0.23  
& 0.24 $\pm$ 0.14  
& 0.19 $\pm$ 0.12  
& 0.38 $\pm$ 0.19\pmark \\

&  H-Learner
&  0.18 $\pm$ 0.08
&  0.23 $\pm$ 0.08\pmark
&  0.19 $\pm$ 0.06\pmark
&  \fontseries{b}\selectfont{0.26 $\pm$ 0.13}
&  \fontseries{b}\selectfont{0.20 $\pm$ 0.10}
&  \fontseries{b}\selectfont{0.18 $\pm$ 0.10}
&  0.31 $\pm$ 0.13\pmark \\

& \fontseries{b}\selectfont{CISI-Net} 
& \fontseries{b}\selectfont{0.16 $\pm$ 0.13}
& \fontseries{b}\selectfont{0.19 $\pm$ 0.13}
& \fontseries{b}\selectfont{0.12 $\pm$ 0.09}
& \fontseries{b}\selectfont{0.26 $\pm$ 0.18}
& 0.23 $\pm$ 0.14
& \fontseries{b}\selectfont{0.18 $\pm$ 0.12}
& \fontseries{b}\selectfont{0.21 $\pm$ 0.14} \\
\hline
\multirow{6}{*}{\centering 3}
& TARNet  
& 0.11 $\pm$ 0.08 
& 0.11 $\pm$ 0.07 
& 0.10 $\pm$ 0.07 
& 0.17 $\pm$ 0.12\pmark
& 0.14 $\pm$ 0.12 
& 0.16 $\pm$ 0.12\pmark 
& 0.31 $\pm$ 0.26\pmark \\

& CFR-WASS 
& 0.11 $\pm$ 0.08 
& 0.11 $\pm$ 0.09 
& 0.10 $\pm$ 0.07   
& 0.16 $\pm$ 0.13\pmark 
& 0.14 $\pm$ 0.11 
& 0.16 $\pm$ 0.12\pmark 
& 0.28 $\pm$ 0.20\pmark \\

& NCoRE    
& \fontseries{b}\selectfont{0.10 $\pm$ 0.08} 
& 0.10 $\pm$ 0.07 
& 0.09 $\pm$ 0.06 
& 0.17 $\pm$ 0.13 
& 0.16 $\pm$ 0.12 
& 0.15 $\pm$ 0.13  
& 0.19 $\pm$ 0.16\pmark \\

& TECE-VAE 
& 0.12 $\pm$ 0.10 
& 0.12 $\pm$ 0.09  
& 0.10 $\pm$ 0.08  
& 0.13 $\pm$ 0.11 
& \fontseries{b}\selectfont{0.12 $\pm$ 0.09} 
& \fontseries{b}\selectfont{0.10 $\pm$ 0.09} 
& 0.11 $\pm$ 0.09 \\

&  H-Learner
&  \fontseries{b}\selectfont{0.10 $\pm$ 0.08}
&  0.10 $\pm$ 0.08
&  0.09 $\pm$ 0.06
&  0.15 $\pm$ 0.13
&  \fontseries{b}\selectfont{0.12 $\pm$ 0.10}
&  0.13 $\pm$ 0.10
&  0.16 $\pm$ 0.13\pmark \\

& \fontseries{b}\selectfont{CISI-Net} 
& \fontseries{b}\selectfont{0.10 $\pm$ 0.06} 
& \fontseries{b}\selectfont{0.09 $\pm$ 0.08} 
& \fontseries{b}\selectfont{0.08 $\pm$ 0.06} 
& \fontseries{b}\selectfont{0.12 $\pm$ 0.09} 
& \fontseries{b}\selectfont{0.12 $\pm$ 0.08} 
& \fontseries{b}\selectfont{0.10 $\pm$ 0.08} 
& \fontseries{b}\selectfont{0.09 $\pm$ 0.07} \\
\hline
\end{tabular}
}
\end{table*}

When extended to multi-treatment settings, single-treatment architectures like TARNet and CFR-WASS are fundamentally limited by their inability to share information across similar treatment patterns.
Although their original design enables relatively accurate estimation of ASE in some scenarios, these methods treat each treatment pattern independently and construct separate outcome prediction networks for each case.
Because parameter updates occur only for samples that receive the corresponding treatment, they cannot use similarities among similar treatment patterns and thus miss opportunities for parameter sharing.
This structural limitation explains why these methods, although competitive in estimating single treatment effects, are considerably less stable when estimating interaction effects.
These results suggest that naively extending single-treatment architectures is insufficient for robust estimation under multiple concurrent treatments.

NCoRE, which possesses interaction subnetworks updated only in samples receiving multiple treatments, demonstrates the limitations of estimating interaction effects.
Since such samples are typically rare, these subnetworks are poorly trained under data sparsity, which results in large $\epsilon_{\mathrm{AIE}}$.
This instability becomes more severe in simulation scenario 2, because the interaction subnetworks must be learned from imbalanced treatment combinations when only proxy variables are available to the model.
Moreover, its architecture, which constructs outcome and interaction networks independently, prevents any parameter sharing across treatment patterns. 
This design misses a critical opportunity to use treatment similarities, which in turn leads to more unstable estimates.
In simulation scenario 3, these interaction subnetworks become redundant, which in turn limits stable estimation when interaction is absent.

Although TECE-VAE captures interaction treatment effects, it underperforms compared to CISI-Net due to unnecessary latent recovery caused by model misspecification.
In simulation scenario 1, where no latent covariates exist, the model is forced to infer unnecessary latent variables, which results in degraded accuracy in both ASE and AIE estimation. 
In simulation scenario 2, which is a favorable setting for TECE-VAE, TECE-VAE remains competitive for some $\epsilon_{\mathrm{AIE}}$ involving two treatments but still underperforms CISI-Net in most metrics, especially in $\epsilon_{\mathrm{AIE}}(\{1,2,3\})$.
These results show that reliance on latent assumptions of TECE-VAE degrades robustness across diverse settings compared to CISI-Net.

H-Learner, which also promotes parameter sharing across treatment patterns, achieves competitive performance for several $\epsilon_{\mathrm{ASE}}$ and $\epsilon_{\mathrm{AIE}}$.
However, CISI-Net achieves lower or comparable estimation errors in most settings and shows improvements in $\epsilon_{\mathrm{AIE}}(\{1,2,3\})$.
These results suggest that the performance of CISI-Net cannot be explained solely by the use of a parameter-sharing mechanism.

\paragraph{Ablation Study}
Table~\ref{tab:simulation_ablation} shows the results of an ablation study on simulation scenario 1 that evaluates the impact of the task embedding network and the balancing penalty on the performance of CISI-Net.
These results demonstrate that whereas each component individually contributes to performance, combining the task embedding network with a balancing penalty yields the lowest estimation errors for all metrics.
The balancing penalty, which reduces estimation error, alone improves AIE estimation because units exposed to multiple treatments are subject to strong selection bias.
The task embedding network captures treatment-effect-related structure among treatment patterns, including similarities in both single and interaction treatment effects, and this mechanism contributes to the accurate estimation of both types of effects.
The interaction of each component enables the model to consistently achieve lower errors across both single and interaction effects compared to using either component in isolation.
The additional results of the ablation study using other simulation datasets are presented in Appendix~\ref{appendix:additional_ablation_results}.

\begin{table*}[tb]
\caption{
Ablation study on simulation dataset 1. 
Here, $k \in \{1,2,3\}$ indexes single treatments for ASE, and $S \subseteq \{1,2,3\}$ (with $|S| \ge 2$) denotes treatment combinations for AIE. 
TE indicates whether the task embedding network is used, and BP indicates whether the balancing penalty is applied. 
When TE is disabled, the original treatment vector is directly concatenated with the learned representation and passed to the shared outcome network.
$\alpha$ is set to 0.1.}
\label{tab:simulation_ablation}
\centering
\setlength{\tabcolsep}{6pt}          
\renewcommand{\arraystretch}{1.2}    
\resizebox{\textwidth}{!}{           
\begin{tabular}{ccccc|cccc}
\hline
\multicolumn{2}{l}{} & \multicolumn{3}{c|}{$\epsilon_{\mathrm{ASE}}$} & \multicolumn{4}{c}{$\epsilon_{\mathrm{AIE}}$} \\
\cline{3-5}\cline{6-9}
\shortstack{TE} & BP & $k=1$ & $k=2$ & $k=3$ & $S=\{1,2\}$ & $S=\{2,3\}$ & $S=\{1,3\}$ & $S=\{1,2,3\}$ \\
\hline
\xmarkgray & \xmarkgray & \fontseries{b}\selectfont{0.10 $\pm$ 0.07} & 0.11 $\pm$ 0.07 & 0.10 $\pm$ 0.05 & 0.18 $\pm$ 0.10 & 0.25 $\pm$ 0.14 & 0.25 $\pm$ 0.10 & 0.66 $\pm$ 0.04 \\
\cmark     & \xmarkgray &  0.11 $\pm$ 0.09 & 0.11 $\pm$ 0.08 & 0.09 $\pm$ 0.07 & 0.18 $\pm$ 0.15 & 0.19 $\pm$ 0.15 & 0.21 $\pm$ 0.14 & 0.14 $\pm$ 0.10 \\
\xmarkgray & \cmark     & \fontseries{b}\selectfont{0.10 $\pm$ 0.08} & 0.10 $\pm$ 0.07 & 0.10 $\pm$ 0.06 & 0.16 $\pm$ 0.11 & 0.23 $\pm$ 0.14 & 0.25 $\pm$ 0.13 & 0.59 $\pm$ 0.07 \\
\cmark     & \cmark     & \fontseries{b}\selectfont{0.10 $\pm$ 0.07} & \fontseries{b}\selectfont{0.09 $\pm$ 0.08} & \fontseries{b}\selectfont{0.08 $\pm$ 0.07} & \fontseries{b}\selectfont{0.12 $\pm$ 0.10} & \fontseries{b}\selectfont{0.13 $\pm$ 0.10} & \fontseries{b}\selectfont{0.12 $\pm$ 0.10} & \fontseries{b}\selectfont{0.12 $\pm$ 0.10}\\
\hline

\end{tabular}
} 
\end{table*}

\paragraph{Treatment-Effect Structure in Task Embeddings}
We analyze whether the task embedding reflects similarities among treatment patterns in terms of their true single treatment effects and interaction treatment effects in simulation scenario 1.
Specifically, based on the outcome-generating function in Equation~\eqref{eq:dgp_outcome_generating_function}, we compute the single treatment effect component for each treatment pattern $\boldsymbol{t}$ and covariate sample $\boldsymbol{x}$ as $f(\boldsymbol{x},\boldsymbol{t},0)-f(\boldsymbol{x},\boldsymbol{0},0)$.
We similarly compute the interaction treatment effect component as $f(\boldsymbol{x},\boldsymbol{t},1)-f(\boldsymbol{x},\boldsymbol{t},0)$.
For each pair of treatment patterns $\boldsymbol{t}$ and $\boldsymbol{t}'$, we compute the single and interaction treatment effect components over the same set of covariate samples and define their root mean squared differences across these samples as $D_S(\boldsymbol{t},\boldsymbol{t}')$ and $D_I(\boldsymbol{t},\boldsymbol{t}')$, respectively.
Accordingly, smaller $D_S$ and $D_I$ indicate that two treatment patterns have more similar single treatment effects and interaction treatment effects, respectively.
We compare these effect-based distances with the Euclidean distances between the same treatment-pattern pairs in (i) the raw treatment-vector space, (ii) the untrained task-embedding space, and (iii) the trained task-embedding space by using Spearman correlations.
This comparison allows us to assess whether the correspondence with the true treatment effects is inherent in the original treatment vectors, present in the untrained task-embedding space, or acquired through training.

\begin{figure}[tb]
  \centering
  \includegraphics[width=\linewidth,  keepaspectratio]{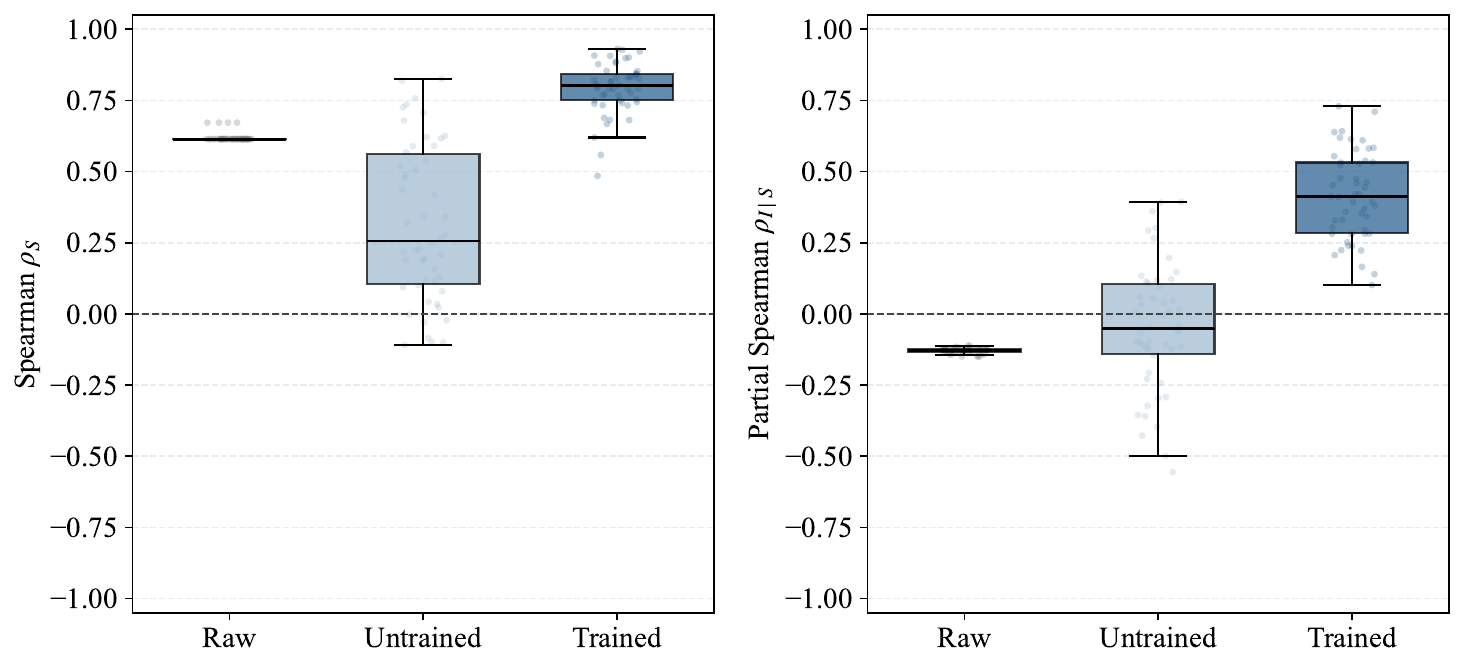}
  \caption{
  Correspondence between treatment-effect distances and distances in the raw treatment-vector and task-embedding spaces in simulation scenario 1.
The left panel shows the Spearman correlation $\rho_S$ between the Euclidean distances in these spaces and $D_S$, and the right panel shows the partial Spearman correlation $\rho_{I\mid S}$ between the Euclidean distances in these spaces and $D_I$ while controlling for $D_S$.
}
  \label{fig:task_embedding_geometry}
\end{figure}

Figure~\ref{fig:task_embedding_geometry} reports the correspondence between the distances in the raw treatment-vector and task-embedding spaces and the true treatment-effect distances.
The partial Spearman correlation $\rho_{I\mid S}$ assesses the correspondence with interaction treatment effects beyond that explained by similarities in single treatment effects.
The trained task embeddings show a stronger correspondence with the single treatment effects than both the raw treatment vectors and the untrained task embeddings. 
Importantly, the trained task embeddings show a clear positive association with similarities in the interaction treatment effects after controlling for similarities in the single treatment effects, whereas the raw treatment vectors show a slightly negative partial correlation and the untrained task embeddings show no consistent association.
These results indicate that training enables the task embedding to acquire treatment-effect-related similarities among treatment patterns.
Together with the ablation results in Table~\ref{tab:simulation_ablation}, these findings support the role of the task embedding network in improving the estimation of both single and interaction treatment effects.

\begin{figure}[tb]
  \centering
  \includegraphics[width=0.75\linewidth, keepaspectratio]{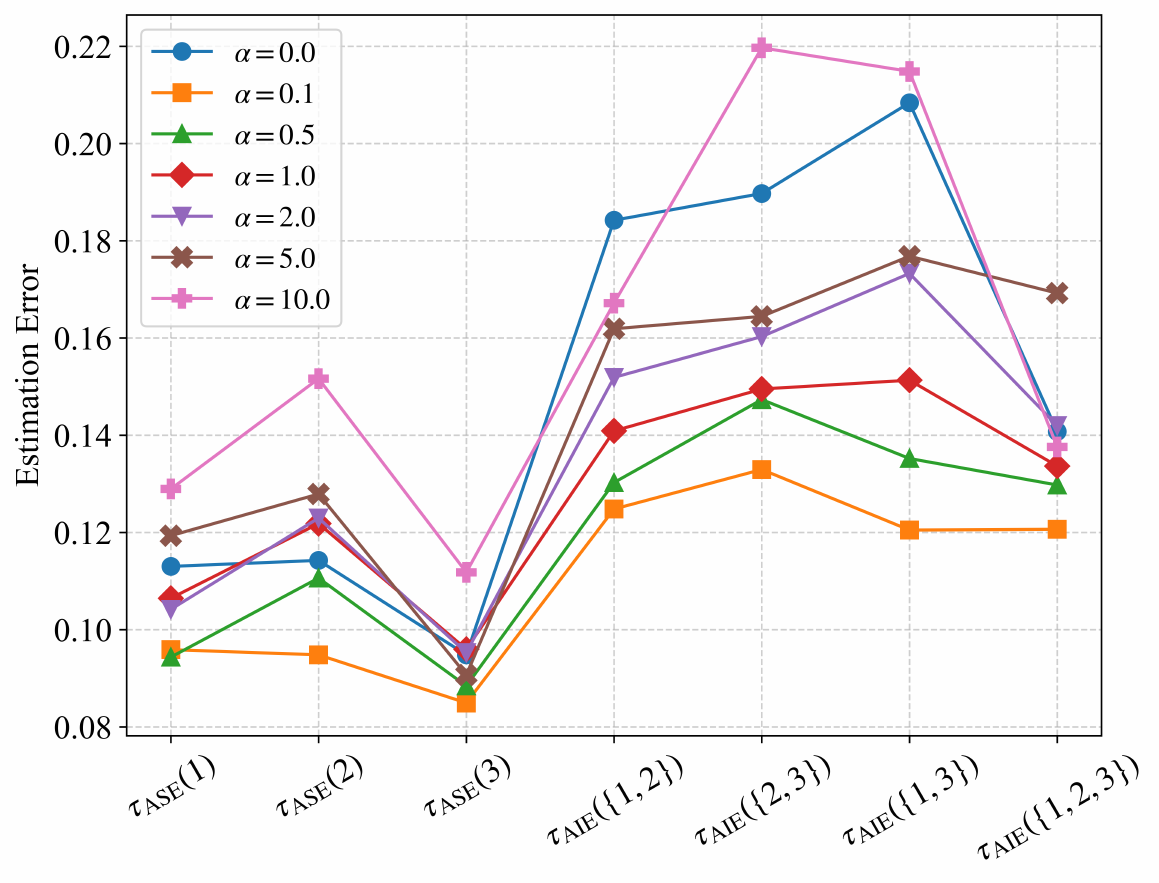}
  \caption{$\epsilon_{\mathrm{ASE}}$ and $\epsilon_{\mathrm{AIE}}$ for different values of the balancing penalty coefficient $\alpha$ in CISI-Net.}
  \label{fig:alpha_sensitivity}
\end{figure}

\paragraph{Sensitivity Analysis}
Figure~\ref{fig:alpha_sensitivity} shows $\epsilon_{\mathrm{ASE}}$ and $\epsilon_{\mathrm{AIE}}$ for different values of the balancing penalty coefficient $\alpha$ of our CISI-Net.
The results reveal that a small positive value for $\alpha$ consistently achieves the lowest $\epsilon_{\mathrm{ASE}}$ and $\epsilon_{\mathrm{AIE}}$, emphasizing the importance of moderate distributional balance.
This finding illustrates a fundamental trade-off inherent to balancing-based causal inference methods.
When $\alpha=0$, although the model prioritizes predictive accuracy, the lack of a balancing penalty introduces higher estimation errors due to distributional imbalances, especially for AIE.
Conversely, an excessively large $\alpha$~(e.g., $\alpha=10$) degrades performance by forcing distributional consistency at the expense of the predictive power of the representation.
This sensitivity is not a limitation unique to our CISI-Net and is a well-documented characteristic of models that employ balancing regularization~\citep{johansson2022generalization,shalit2017estimating}.
Therefore, developing a mechanism to adjust $\alpha$ based on data characteristics adaptively presents a promising direction for future research, which could enhance the stability and versatility of this entire class of models.

\begin{figure*}[tb]
  \centering
  \includegraphics[width=\linewidth, height=0.4\textheight, keepaspectratio]{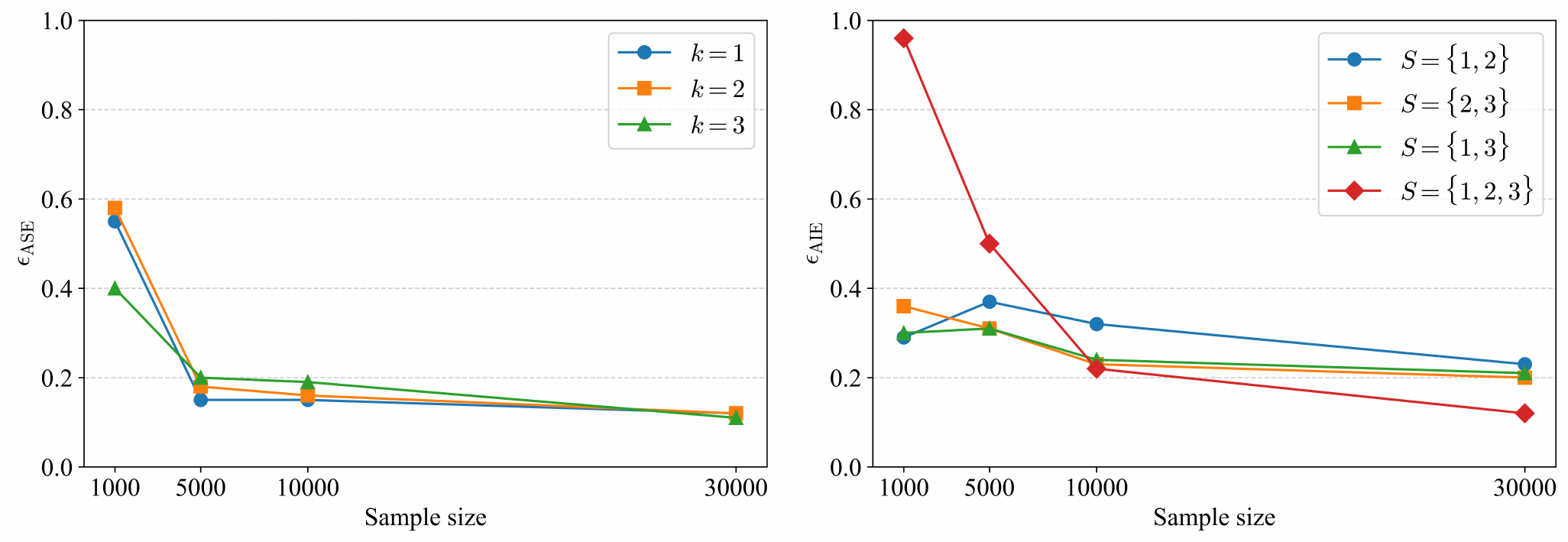}
  \caption{Estimation errors of CISI-Net under varying sample sizes on simulation dataset 1.
  The left panel shows $\epsilon_{\mathrm{ASE}}$, and the right panel shows $\epsilon_{\mathrm{AIE}}$.
  Each line represents the mean error computed over 100 random seeds.}
  \label{fig:sample_error_comparison}
\end{figure*}

\paragraph{Sample Size Analysis}
Figure~\ref{fig:sample_error_comparison} shows $\epsilon_{\mathrm{ASE}}$ and $\epsilon_{\mathrm{AIE}}$ of CISI-Net under different sample sizes.
The results indicate that both $\epsilon_{\mathrm{ASE}}$ and $\epsilon_{\mathrm{AIE}}$ decrease as the sample size increases, but $\epsilon_{\mathrm{AIE}}$ remain high even at $N=10000$.
In the multiple-treatment setting, this strict sample requirement for estimating causal effects is a universal challenge because the number of available samples for each treatment pattern decreases as the number of treatment combinations increases~\citep{Chauhan_2025,qian2021estimating}.
A possible future direction is to integrate complementary data augmentation methods, which follow a different paradigm, into CISI-Net.
In particular, CISI-Net can be extended with a data augmentation strategy that uses learned task embeddings to identify related treatment combinations and generate informative synthetic samples for estimating interaction effects for rare treatment patterns.

\paragraph{Computational Scalability}
Although CISI-Net maintains its estimation performance as the number of treatments increases~(See Appendix~\ref{appendix:scalability_treatment_num_exp}), pairwise balancing incurs increasing computational cost as the number of treatment patterns grows.
The runtime analysis in Appendix~\ref{appendix:computing_cost} shows that the training time increases from 14.90~s/epoch at $K=3$ to 2229.83~s/epoch at $K=10$ under the same hardware and training configuration.
These results indicate that computational scalability remains a limitation of CISI-Net when the number of treatments becomes large.
\section{Application to Multiple Marketing Promotions}
\label{sec:real_world_application}
In this section, we apply CISI-Net to a real-world dataset derived from multiple marketing promotions.
The marketing promotions were conducted by merchants affiliated with a mobile payment platform operated by a mobile payment provider in Japan. 
The objective of this analysis is to examine the practical application of CISI-Net to estimating both single and interaction treatment effects from observational data.

We use two real-world marketing datasets in which multiple marketing promotions were simultaneously conducted.
The first dataset includes three marketing promotions: two promotions organized by the same merchant group~(denoted as CP\textsubscript{1} and CP\textsubscript{2}) and one online campaign conducted by another merchant group~(denoted as CP\textsubscript{3}). 
The observed sample proportions were approximately 2\% for CP\textsubscript{1} only, 15\% for CP\textsubscript{2} only, 42\% for CP\textsubscript{3} only, 0.2–3\% for combinations of two CPs, 0.2\% for all three promotions simultaneously, and the remainder as the control group.
The second dataset includes two marketing promotions conducted by two different merchants in the same industry~(denoted as CP\textsubscript{4} and CP\textsubscript{5}).
The observed sample proportions in the second dataset were approximately 29\% for CP\textsubscript{4} only, 5\% for CP\textsubscript{5} only, 1\% for both promotions, and the remainder as the control group.
In all datasets, the control group consisted of users who were not exposed to any promotion during the treatment period and were randomly sampled from those who had at least one mobile payment transaction in the month before the promotions.

In all datasets, we defined the outcome as the total payment amount during the one month following campaign implementation, and empirically analyzed the causal effects of the two datasets.
For numerical stability and confidentiality, the outcome $Y$ was standardized within each dataset by using its own overall mean and standard deviation.
Accordingly, the results reported in this section are based on standardized outcomes.
The original-scale causal effects can be recovered by applying the inverse transformation of the scaler to the model outputs. 
This preprocessing does not distort the estimation results because standardization is an affine transformation~\citep{thakral2023estimates}. 
In particular, consistent interpretation between the standardized and original scales is ensured because the signs of the estimated causal effects are preserved.
The observed covariates consisted of 71 variables, which included service usage histories and user demographic attributes.

The hyperparameters of the proposed model were fixed to the same values as those described in Section~\ref{subsec:implement_details}. 
A sensitivity analysis regarding the balancing-penalty weight $\alpha$ in Equation~\eqref{eq:proposed_loss} is presented in Appendix~\ref{appendix:sensitivity_analysis_in_empirical_applications}.
Each dataset is divided into training sets of 70\% and test sets of 30\%, and causal effects were estimated on the test set according to Equations \eqref{ASE} and \eqref{AIE}.

To quantify uncertainty, we performed 1,000 nonparametric bootstrap replications for a fixed hyperparameter configuration~\citep{efron1994introduction,kubota2025impact}.
In each replication, we resample users with replacement and retrain CISI-Net under the same hyperparameter configuration.
We estimate each ASE and AIE as the mean of the corresponding unit-level effect estimates over the test set.
For each effect and configuration, we construct a 95\% percentile bootstrap confidence interval from the empirical distribution of the 1,000 replicate-level estimates.

\begin{figure*}[tb]
  \centering
  \includegraphics[width=\linewidth,  keepaspectratio]{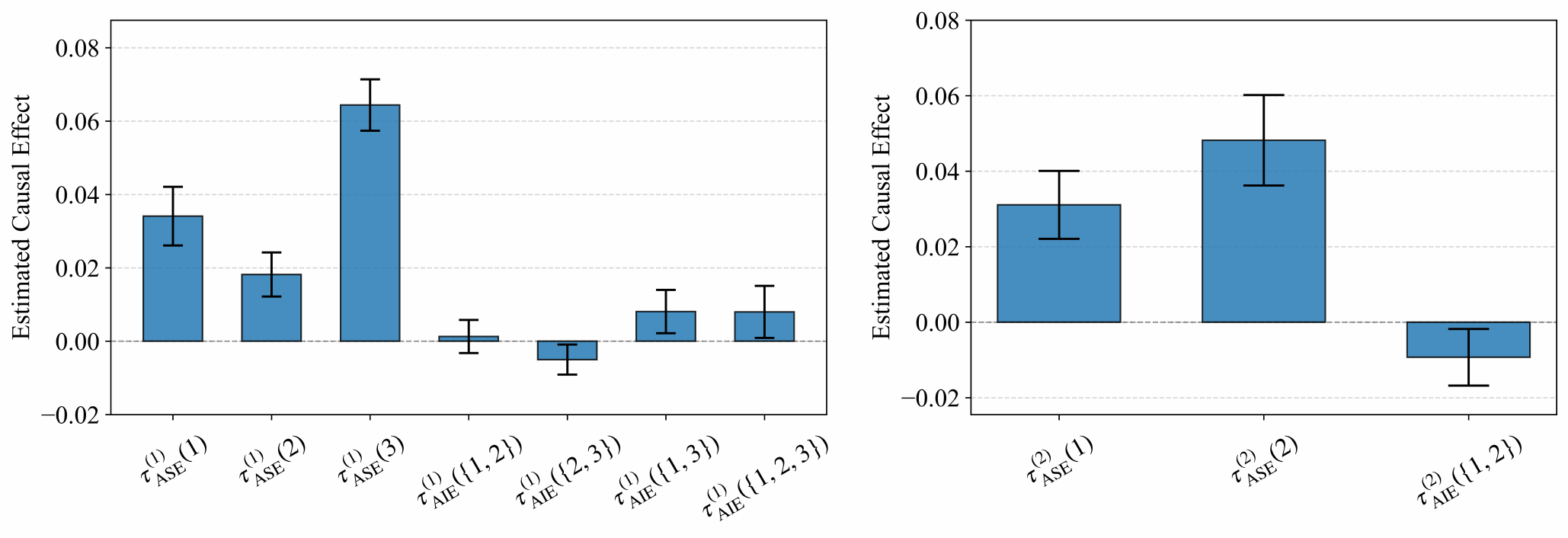}
  \caption{Estimated single and interaction treatment effects obtained from two real-world marketing promotion datasets. 
  The left panel corresponds to dataset 1~(CP\textsubscript{1}-CP\textsubscript{3}), and the right panel corresponds to dataset 2~(CP\textsubscript{4} and CP\textsubscript{5}).
  Here, $\smash{\tau_{\mathrm{ASE}}^{(d)}(k)}$ and $\smash{\tau_{\mathrm{AIE}}^{(d)}(S)}$ are the estimated effects for single and multiple treatments in dataset $d$.
  The outcome is standardized prior to estimation.
  Bars show point estimates, and error bars indicate 95\% bootstrap confidence intervals based on 1,000 replications.}
  \label{fig:empirical_result}
\end{figure*}

Figure~\ref{fig:empirical_result} shows the estimated causal effects obtained from two real-world marketing promotion datasets on a standardized outcome scale.
In the first dataset~(left panel), all estimated single treatment effects were positive, with CP\textsubscript{3} achieving the largest single treatment effect among the three promotions.
The point estimate of the interaction treatment effect $\smash{\tau_{\mathrm{AIE}}^{(1)}(\{1,2\})}$ between two promotions conducted by the same merchant group was positive, but the evidence for a positive interaction was inconclusive because the corresponding bootstrap confidence interval included zero.
On the other hand, the estimated interaction treatment effects between promotions conducted by different merchant groups~($\smash{\tau_{\mathrm{AIE}}^{(1)}(\{2,3\})}$, $\smash{\tau_{\mathrm{AIE}}^{(1)}(\{1,3\})}$, and $\smash{\tau_{\mathrm{AIE}}^{(1)}(\{1,2,3\})}$) exhibited heterogeneous patterns.
In the second dataset~(right panel), both estimated single treatment effects were positive, and their 95\% bootstrap confidence intervals were entirely above zero.
The interaction effect $\smash{\tau_{\mathrm{AIE}}^{(2)}(\{1,2\})}$ was negative and its bootstrap confidence interval lay entirely below zero, which implies cannibalization between similar promotions conducted by different merchants.

The results from the first dataset suggest that the direction and magnitude of the interaction treatment effect vary depending on the combination of promotions.
The heterogeneous patterns of interaction effects observed for cross-group promotions~(e.g., $\smash{\tau_{\mathrm{AIE}}^{(1)}(\{2,3\})}$ and $\smash{\tau_{\mathrm{AIE}}^{(1)}(\{1,3\})}$) indicate that promotional effects may vary depending on the degree of market overlap between different merchant groups.
These results are consistent with findings from partnership loyalty programs, where collaboration between distinct partners can result in either synergy or cannibalization depending on context~\citep{dorotic2021synergistic}.

The results from the second dataset suggest that cannibalization occurs when competing merchants conduct promotions simultaneously.
One possible interpretation is that consumers split their spending across merchants when similar promotions are conducted simultaneously in the same industry, which may limit the combined promotional effect relative to the sum of the single effects.
Our analysis is consistent with previous research indicating that overlapping promotions from competing merchants tend to create cannibalization rather than synergy~\citep{dorotic2021synergistic}.
Because the validity of these estimated causal effects depends on the identification assumptions, violations of these assumptions may lead to inaccurate estimates and inappropriate marketing decisions.
Therefore, the results should be interpreted together with the diagnostics reported in Appendix~\ref{appendix:real_world_diagnostics}.
While these considerations are important for interpreting the results, overall, these results show the applicability of CISI-Net to estimating single and interaction treatment effects across promotion combinations, with the estimated patterns being qualitatively consistent with empirically validated behavioral patterns.

\section{Conclusion}
In this study, we propose the Causal Inference for Single and Interaction treatment effects Network~(CISI-Net), a novel deep learning framework specifically designed to estimate both single and interaction effects in multi-treatment scenarios.
CISI-Net estimates causal effects by combining a task embedding network with a representation learning network with the balancing penalty.
CISI-Net facilitates parameter sharing across treatment patterns by using the task embedding network, which learns treatment effect-related similarities among treatment patterns, including interaction-related structures, thereby supporting accurate estimation of both single and interaction treatment effects.
Additionally, CISI-Net reduces distributional discrepancies across treatment patterns by using the representation learning network with the balancing penalty, thereby controlling a source of estimation error in causal effect estimation.

The experimental results demonstrate the effectiveness of CISI-Net.
First, the simulation studies show that CISI-Net achieves lower estimation errors than existing baselines in most settings for both single and interaction treatment effects across a wide range of conditions.
Second, the real-world case study demonstrates the practical utility of CISI-Net by estimating both single and interaction effects from observational promotional data, with the estimated patterns being qualitatively consistent with prior marketing research.
Because the interpretation of estimated causal effects in observational studies depends on identification assumptions, CISI-Net should be applied with appropriate diagnostics and careful consideration of these assumptions.
Subject to these considerations, these results suggest that CISI-Net has great potential as a practical analytical tool for applications such as evaluating combined drug effects in medicine and optimizing complex promotional strategies in marketing.

Our work opens up several important avenues for future research.
A first direction is to further improve the balancing component in terms of both adaptive adjustment of the weight $\alpha$ of the IPM-based balancing penalty and computational scalability.
A second direction is to investigate methodological extensions to address the strict sample requirements. 
This direction particularly indicates investigating methods for stably estimating interaction treatment effects, even under conditions where sample sizes per treatment pattern are severely limited, such as those involving simultaneous multiple treatments.
Possible directions for future research include incorporating data augmentation–based approaches to stabilize parameter estimation.
A third direction is to further reduce the risk of model misspecification by extending our framework to a doubly robust estimation setting~\citep{kennedy2023towards} that integrates both propensity score modeling and outcome regression.

\bibliographystyle{tmlr}

\appendix
\section*{Appendix}
\section{Proofs of Identifiability}
\label{appendix:proofs_of_identifiability}
This appendix provides the proofs for the identifiability of the causal estimands introduced in Section~\ref{sec:causal-inference}.
We first provide the proof of Proposition 1.
\begin{proof}
The proof proceeds by a sequence of equalities, starting from the definition of $\mu(\boldsymbol{x}, \boldsymbol{t})$:
\begin{align*}
\mu(\boldsymbol{x}, \boldsymbol{t}) & \coloneqq  \mathbb{E}[Y(\boldsymbol{t}) \mid \boldsymbol{X} = \boldsymbol{x}] && \text{(by Definition in~(1))} \\
&= \mathbb{E}[Y(\boldsymbol{t}) \mid \boldsymbol{X} = \boldsymbol{x}, \boldsymbol{T} = \boldsymbol{t}] && \text{(by Assumption 2)} \\
&= \mathbb{E}[Y \mid \boldsymbol{X} = \boldsymbol{x}, \boldsymbol{T} = \boldsymbol{t}] && \text{(by Assumption 1)}
\end{align*}
Assumption 3 ensures this final quantity is well-defined.
This completes the proof of identifiability for $\mu(\boldsymbol{x}, \boldsymbol{t})$.
\end{proof}

Second, we provide the proof of Corollary 1.
\begin{proof}
The ASE and AIE are defined as functions of the conditional average potential outcome $\mu(\boldsymbol{x}, \boldsymbol{t})$.
Specifically, they are constructed using linear combinations and expectations over the distributions of $\boldsymbol{X}$.
Since Proposition 1 establishes that $\mu(\boldsymbol{x}, \boldsymbol{t})$ is identifiable from observed data, any quantities derived from it through these operations are also identifiable.
\end{proof}

\section{Estimation-Error Bounds for Pairwise IPM Balancing}
\label{appendix:proof_upper_bounds}
\subsection{Setup and Definitions}
\label{app:cisi:setup}

\paragraph{Problem setup.}
Let $\mathcal{X}\subseteq\mathbb{R}^{d}$ be the covariate space and let $\mathcal{Y}\subseteq\mathbb{R}$ be the outcome space.
Let $\mathcal{T}=\{0,1\}^{K}$ be the set of treatment patterns for $K\geq 1$ binary treatments.
The random variables $\boldsymbol{X}\in\mathcal{X}$ and $\boldsymbol{T}\in\mathcal{T}$ have realizations $\boldsymbol{x}$ and $\boldsymbol{t}$, respectively.
For every $\boldsymbol{t}\in\mathcal{T}$, let $Y(\boldsymbol{t})\in\mathcal{Y}$ denote the potential outcome under treatment pattern $\boldsymbol{t}$, and let $Y$ denote the observed outcome.
We write $p(\boldsymbol x)$ for the marginal covariate distribution,
$p(\boldsymbol t) \coloneqq \Pr(\boldsymbol T=\boldsymbol t)$ for the marginal treatment-pattern probability,
and $p(\boldsymbol x\mid\boldsymbol t)$ for the covariate distribution conditional on $\boldsymbol T=\boldsymbol t$.
Throughout this appendix, all distributions and expected losses are considered at the population level.
The plug-in estimates of average causal effects are defined by integration over $p(\boldsymbol x)$.

Let $\Phi:\mathcal{X}\to\mathcal{R}\subseteq\mathbb{R}^{p}$ be the measurable representation map, where $p$ is the representation dimension and $\mathcal{R}=\Phi(\mathcal{X})$ is its image.
Let $t_w:\mathcal{T}\to\mathbb{R}^{q}$ be the task embedding map, and let $h:\mathbb{R}^{p+q}\to\mathbb{R}$ be the outcome prediction function.
The predictor is $\hat\mu(\boldsymbol{x},\boldsymbol{t}) \coloneqq h\!\left([\Phi(\boldsymbol{x}),t_w(\boldsymbol{t})]\right)$, where $[\cdot,\cdot]$ denotes concatenation.
For each $\boldsymbol t\in\mathcal T$, let $R_{\boldsymbol t}$ denote the distribution of $\Phi(\boldsymbol X)$ conditional on $\boldsymbol T=\boldsymbol t$.
Thus, $R_{\boldsymbol t}$ is the pushforward of $p(\boldsymbol x\mid\boldsymbol t)$ under $\Phi$.

\begin{definition}
\label{def:cisi:losses}
For covariates $\boldsymbol{x}\in\mathcal{X}$ and treatment pattern $\boldsymbol{t}\in\mathcal{T}$, define the conditional expected squared loss as $ l(\boldsymbol{x},\boldsymbol{t}) \coloneqq  \int \left( Y(\boldsymbol{t})-\hat{\mu}(\boldsymbol{x},\boldsymbol{t}) \right)^2 p\!\left(Y(\boldsymbol{t})\mid\boldsymbol{x}\right) \,dY(\boldsymbol{t}). $
The dependence of $l$ on $(\Phi,t_w,h)$ is suppressed in the notation.
The expected factual and counterfactual losses for treatment pattern $\boldsymbol{t}$ are defined as
\begin{align}
\epsilon_{\mathrm F}^{(\boldsymbol{t})}
& \coloneqq p(\boldsymbol{t})
\int_{\mathcal{X}}
l(\boldsymbol{x},\boldsymbol{t})
p(\boldsymbol{x}\mid\boldsymbol{t})
\,d\boldsymbol{x},
\label{app:cisi:factual_loss}\\
\epsilon_{\mathrm{CF}}^{(\boldsymbol{t})}
& \coloneqq 
\sum_{\substack{\boldsymbol{s}\in\mathcal{T}\\
\boldsymbol{s}\ne\boldsymbol{t}}}
p(\boldsymbol{s})
\int_{\mathcal{X}}
l(\boldsymbol{x},\boldsymbol{t})
p(\boldsymbol{x}\mid\boldsymbol{s})
\,d\boldsymbol{x}.
\label{app:cisi:counterfactual_loss}
\end{align}
The second quantity concerns prediction under treatment pattern $\boldsymbol{t}$ for units whose observed patterns differ from $\boldsymbol{t}$.
\end{definition}
Our notation $\epsilon_{\mathrm F}^{(\boldsymbol{t})}$ includes the factor $p(\boldsymbol{t})$.
Its ratio to $p(\boldsymbol{t})$ is the treatment-conditional factual risk.
This distinction is necessary when this notation is compared with treatment-conditional factual losses in \citet{shalit2017estimating}.

\begin{definition}
\label{def:cisi:ipm}
Let $\mathcal{F}$ be a nonempty class of measurable real-valued functions on $\mathcal{R}$.
For probability measures $P$ and $Q$ on $\mathcal{R}$, define
\begin{equation*}
\operatorname{IPM}_{\mathcal{F}}(P,Q)
 \coloneqq 
\sup_{f\in\mathcal{F}}
\left|
\int_{\mathcal{R}}f(\boldsymbol r)\,dP(\boldsymbol r)
-
\int_{\mathcal{R}}f(\boldsymbol r)\,dQ(\boldsymbol r)
\right|.
\end{equation*}
\end{definition}
We use the same function class for all treatment-pattern pairs.
All function expectations and all IPMs considered below are assumed finite.
For notational convenience, we write
\begin{equation*}
D_{\boldsymbol{t},\boldsymbol{s}}
 \coloneqq \operatorname{IPM}_{\mathcal F}(R_{\boldsymbol t},R_{\boldsymbol s}).
\end{equation*}
The definition implies $D_{\boldsymbol{t}, \boldsymbol{s}}=D_{\boldsymbol{s}, \boldsymbol{t}}\geq0$
and $D_{\boldsymbol{t}, \boldsymbol{t}}=0$.

\begin{definition}[Pairwise balancing discrepancy]
\label{def:cisi:population_penalty}
Let $\binom{\mathcal T}{2}$ denote the set of unordered pairs of distinct treatment patterns, and let
\[
C_{\mathcal T} \coloneqq \binom{|\mathcal T|}{2}=\binom{2^K}{2}.
\]
Define
\begin{equation*}
L_{\Phi}
 \coloneqq \frac{1}{C_{\mathcal T}}
\sum_{\{\boldsymbol u,\boldsymbol v\}\in\binom{\mathcal T}{2}}
D_{\boldsymbol u,\boldsymbol v}.
\end{equation*}
This quantity is the pairwise IPM balancing penalty evaluated using the treatment-specific representation distributions $R_{\boldsymbol t}$.
\end{definition}

\begin{assumption}[Stable Unit Treatment Value Assumption]
\label{ass:cisi:consistency}
(1) no interference, meaning that the outcome of one unit is unaffected by the treatment assignments of other units; and (2) consistency of treatment, meaning that the potential outcomes correspond to well-defined and unique treatments~(i.e., $Y_i = Y_i(\boldsymbol{T}_i)$).
\end{assumption}

\begin{assumption}[Ignorability]
\label{ass:cisi:ignorability}
For any treatment pattern, the potential outcome is independent of the assigned treatment $T$ given the observed covariates $X$.
Formally, for all $\boldsymbol{t}$, $$ Y(\boldsymbol{t}) \perp \boldsymbol{T} \mid \boldsymbol{X} $$
\end{assumption}

\begin{assumption}[Overlap]
\label{ass:cisi:overlap}
Every unit has a non-zero probability of receiving any treatment pattern given its observed covariates.
Formally, for all $\boldsymbol{t}$ and $\boldsymbol{x}$, $$ 0 < P(\boldsymbol{T}=\boldsymbol{t} \mid \boldsymbol{X}=\boldsymbol{x}) < 1 $$
\end{assumption}

\begin{assumption}[Invertible representation]
\label{ass:cisi:invertibility}
The map $\Phi$ is one-to-one on $\mathcal{X}$ and admits a measurable inverse $\Psi:\mathcal{R}\to\mathcal{X}$ such that
\[
\Psi(\Phi(\boldsymbol{x}))=\boldsymbol{x}
\qquad\text{for all }\boldsymbol{x}\in\mathcal{X}.
\]
\end{assumption}

\begin{assumption}[Compatibility of the expected loss with the IPM class]
\label{ass:cisi:loss_class}
For the fixed predictor and representation, there exists a finite constant $B_{\Phi}>0$ such that
\begin{equation*}
g_{\boldsymbol{t}}(\boldsymbol r)
 \coloneqq \frac{l(\Psi(\boldsymbol r),\boldsymbol{t})}{B_{\Phi}}
\in\mathcal F
\qquad\text{for any }\boldsymbol{t}\in\mathcal T.
\end{equation*}
The constant is common to all treatment patterns.
\end{assumption}

This is the loss-class inclusion condition used in the general IPM argument of \citet{shalit2017estimating}, now imposed for every treatment pattern.

\subsection{A General IPM Bound for Counterfactual Prediction}
\label{app:cisi:ipm_prediction_bounds}
To connect representation discrepancies with counterfactual prediction error, we first derive an upper bound that relates prediction losses across different treatment-specific covariate distributions through the IPM between their representation distributions.
We then aggregate these pairwise comparisons over the treatment patterns to bound the counterfactual loss for each treatment pattern by its factual prediction loss and the corresponding pairwise IPM discrepancies.

\begin{lemma}
\label{lem:cisi:source_target}
Under Assumptions~\ref{ass:cisi:invertibility} and~\ref{ass:cisi:loss_class}, for any $\boldsymbol t,\boldsymbol s\in\mathcal T$,
\begin{equation*}
        \int_{\mathcal{X}}l(\boldsymbol{x}, \boldsymbol{t})p(\boldsymbol{x}\mid \boldsymbol{s})d\boldsymbol{x} \leq \int_{\mathcal{X}}l(\boldsymbol{x}, \boldsymbol{t})p(\boldsymbol{x}\mid\boldsymbol{t})d\boldsymbol{x} + B_{\Phi} D_{\boldsymbol{t}, \boldsymbol{s}}.
    \end{equation*}
\end{lemma}
\begin{proof}
By the definition of the representation distributions and
Assumption~\ref{ass:cisi:invertibility}, we have
\begin{align*}
\int_{\mathcal X}l(\boldsymbol x,\boldsymbol t)p(\boldsymbol x\mid\boldsymbol s)\,d\boldsymbol x
&=\int_{\mathcal R}l(\Psi(\boldsymbol r),\boldsymbol t)\,dR_{\boldsymbol s}(\boldsymbol r),\\
\int_{\mathcal X}l(\boldsymbol x,\boldsymbol t)p(\boldsymbol x\mid\boldsymbol t)\,d\boldsymbol x
&=\int_{\mathcal R}l(\Psi(\boldsymbol r),\boldsymbol t)\,dR_{\boldsymbol t}(\boldsymbol r).
\end{align*}
Therefore, the difference between the two expected losses satisfies
\begin{align*}
&\int_{\mathcal X}l(\boldsymbol x,\boldsymbol t)p(\boldsymbol x\mid\boldsymbol s)\,d\boldsymbol x
- \int_{\mathcal X}l(\boldsymbol x,\boldsymbol t)p(\boldsymbol x\mid\boldsymbol t)\,d\boldsymbol x\\
&=\int_{\mathcal R}l(\Psi(\boldsymbol r),\boldsymbol t)\bigl(dR_{\boldsymbol s}(\boldsymbol r)
-dR_{\boldsymbol t}(\boldsymbol r)\bigr)\\
&=B_{\Phi}\int_{\mathcal R}g_{\boldsymbol t}(\boldsymbol r)\bigl(dR_{\boldsymbol s}(\boldsymbol r)
-dR_{\boldsymbol t}(\boldsymbol r)\bigr)\\
&\leq B_{\Phi} \left|\int_{\mathcal R}g_{\boldsymbol t}(\boldsymbol r)\bigl(dR_{\boldsymbol s}(\boldsymbol r)
-dR_{\boldsymbol t}(\boldsymbol r)\bigr)\right|\\
&\leq B_{\Phi} \sup_{f\in\mathcal F}\left|\int_{\mathcal R}f(\boldsymbol r)\bigl(dR_{\boldsymbol s}(\boldsymbol r) - dR_{\boldsymbol t}(\boldsymbol r)\bigr)\right|\\
&=B_{\Phi} \operatorname{IPM}_{\mathcal F} (R_{\boldsymbol s},R_{\boldsymbol t})
= B_{\Phi}D_{\boldsymbol t,\boldsymbol s}.
\end{align*}
The first inequality follows from $z\leq |z|$ for any real $z$
and $B_{\Phi}>0$.
The second inequality follows from
$g_{\boldsymbol t}\in\mathcal F$ under
Assumption~\ref{ass:cisi:loss_class}.
The last equality uses the symmetry of the IPM.
Consequently,
\[
\int_{\mathcal X}
l(\boldsymbol x,\boldsymbol t)
p(\boldsymbol x\mid\boldsymbol s)\,d\boldsymbol x
\leq
\int_{\mathcal X}
l(\boldsymbol x,\boldsymbol t)
p(\boldsymbol x\mid\boldsymbol t)\,d\boldsymbol x
+
B_{\Phi}D_{\boldsymbol t,\boldsymbol s},
\]
which proves the result.
\end{proof}

\begin{lemma}
\label{lem:cisi:cf_bound}
Under Assumptions~\ref{ass:cisi:consistency}--\ref{ass:cisi:loss_class}, for any $\boldsymbol t\in\mathcal T$,
\begin{equation*}
\epsilon_{\mathrm{CF}}^{(\boldsymbol t)}
\leq
\left(\frac{1}{p(\boldsymbol t)}-1\right)
\epsilon_{\mathrm F}^{(\boldsymbol t)}
+
B_{\Phi}
\sum_{\substack{\boldsymbol s\in\mathcal T\\\boldsymbol s\ne\boldsymbol t}}
p(\boldsymbol s)D_{\boldsymbol t,\boldsymbol s}.
\end{equation*}
\end{lemma}
\begin{proof}
Lemma~\ref{lem:cisi:source_target} implies, for each $\boldsymbol s\ne\boldsymbol t$,
\[
\int_{\mathcal X}l(\boldsymbol x,\boldsymbol t)
 p(\boldsymbol x\mid\boldsymbol s)\,d\boldsymbol x
\leq
\frac{\epsilon_{\mathrm F}^{(\boldsymbol t)}}{p(\boldsymbol t)}
+B_{\Phi}D_{\boldsymbol t,\boldsymbol s}.
\]
We multiply this inequality by $p(\boldsymbol s)$ and sum over $\boldsymbol s\ne\boldsymbol t$.
By Definition~\ref{def:cisi:losses},
\begin{align*}
\epsilon_{\mathrm{CF}}^{(\boldsymbol t)}
&\leq
\left(\sum_{\boldsymbol s\ne\boldsymbol t}p(\boldsymbol s)\right)
\frac{\epsilon_{\mathrm F}^{(\boldsymbol t)}}{p(\boldsymbol t)}
+B_{\Phi}\sum_{\boldsymbol s\ne\boldsymbol t}
p(\boldsymbol s)D_{\boldsymbol t,\boldsymbol s}\\
&=
\left(\frac{1}{p(\boldsymbol t)}-1\right)
\epsilon_{\mathrm F}^{(\boldsymbol t)}
+B_{\Phi}\sum_{\boldsymbol s\ne\boldsymbol t}
p(\boldsymbol s)D_{\boldsymbol t,\boldsymbol s}.
\end{align*}
The treatment-pattern probabilities are retained throughout the argument.
\end{proof}
After we add $\epsilon_{\mathrm F}^{(\boldsymbol t)}$ to both sides of the inequality in Lemma~\ref{lem:cisi:cf_bound}, we obtain
\begin{equation}
\label{app:cisi:marginal_risk_bound}
\epsilon_{\mathrm F}^{(\boldsymbol t)}+
\epsilon_{\mathrm{CF}}^{(\boldsymbol t)}
\leq U_{\boldsymbol t},
\qquad
U_{\boldsymbol t} \coloneqq 
\frac{\epsilon_{\mathrm F}^{(\boldsymbol t)}}{p(\boldsymbol t)}
+
B_{\Phi}
\sum_{\substack{\boldsymbol s\in\mathcal T\\\boldsymbol s\ne\boldsymbol t}}
p(\boldsymbol s)D_{\boldsymbol t,\boldsymbol s}.
\end{equation}

\subsection{Squared Errors of Average Causal Contrasts}
\label{app:cisi:contrast_bounds}
Having bounded the prediction error for each treatment pattern in terms of its factual prediction loss and pairwise IPM discrepancies, we next translate these treatment-specific bounds into an estimation-error bound for average causal contrasts.
To this end, we first express a general causal contrast as a linear combination of treatment-specific average potential outcomes and then relate its squared estimation error to the treatment-specific prediction errors bounded in the previous subsection.
\begin{definition}
\label{def:cisi:contrast}
Let $\boldsymbol a=(a_{\boldsymbol t})_{\boldsymbol t\in\mathcal T}$ be a fixed nonzero vector.
Define its support and squared Euclidean norm by
\[
\mathcal A \coloneqq \{\boldsymbol t\in\mathcal T:a_{\boldsymbol t}\ne0\},
\qquad
\|\boldsymbol a\|_2^2 \coloneqq \sum_{\boldsymbol t\in\mathcal A}a_{\boldsymbol t}^2.
\]
The target and its plug-in estimate are
\begin{align*}
\tau_{\boldsymbol a}
& \coloneqq \sum_{\boldsymbol t\in\mathcal A}a_{\boldsymbol t}
\int_{\mathcal X}\mu(\boldsymbol x,\boldsymbol t)p(\boldsymbol x)\,d\boldsymbol x,
\\
\hat\tau_{\boldsymbol a}
& \coloneqq \sum_{\boldsymbol t\in\mathcal A}a_{\boldsymbol t}
\int_{\mathcal X}\hat\mu(\boldsymbol x,\boldsymbol t)p(\boldsymbol x)\,d\boldsymbol x.
\end{align*}
The contrasts of interest satisfy $\sum_{\boldsymbol t}a_{\boldsymbol t}=0$, although the bound below also holds for a general fixed linear combination.
Define
\begin{equation*}
\mathcal E_{\boldsymbol a}
 \coloneqq \left(\hat\tau_{\boldsymbol a}-\tau_{\boldsymbol a}\right)^2.
\end{equation*}
\end{definition}

\begin{lemma}
\label{lem:cisi:squared_loss}
For each $\boldsymbol t\in\mathcal T$, define
\begin{align*}
\delta_{\boldsymbol t}(\boldsymbol x)
& \coloneqq \hat\mu(\boldsymbol x,\boldsymbol t)-\mu(\boldsymbol x,\boldsymbol t),\\
\mathcal Q_{\boldsymbol t}
& \coloneqq \int_{\mathcal X}\delta_{\boldsymbol t}(\boldsymbol x)^2p(\boldsymbol x)\,d\boldsymbol x,\\
\nu_{\boldsymbol t}
& \coloneqq \mathbb E_{\boldsymbol X}
\left[\operatorname{Var}\!\left(Y(\boldsymbol t)\mid\boldsymbol X\right)\right].
\end{align*}
Under the assumptions of Lemma~\ref{lem:cisi:cf_bound},
\begin{equation}
\label{app:cisi:squared_loss_decomposition}
\mathcal Q_{\boldsymbol t}
=
\epsilon_{\mathrm F}^{(\boldsymbol t)}+
\epsilon_{\mathrm{CF}}^{(\boldsymbol t)}-\nu_{\boldsymbol t}
\leq U_{\boldsymbol t}-\nu_{\boldsymbol t}
\leq U_{\boldsymbol t}.
\end{equation}
\end{lemma}
\begin{proof}
The conditional mean of $Y(\boldsymbol t)-\mu(\boldsymbol x,\boldsymbol t)$ is zero at $\boldsymbol X=\boldsymbol x$.
An expansion of the square in the definition of $l$ (Definition~\ref{def:cisi:losses}) therefore gives
\[
l(\boldsymbol x,\boldsymbol t)
=
\delta_{\boldsymbol t}(\boldsymbol x)^2
+
\operatorname{Var}\!\left(Y(\boldsymbol t)\mid\boldsymbol X=\boldsymbol x\right).
\]
The mixture identity
$p(\boldsymbol x)=\sum_{\boldsymbol s\in\mathcal T}p(\boldsymbol s)p(\boldsymbol x\mid\boldsymbol s)$
and Definition~\ref{def:cisi:losses} imply
\[
\int_{\mathcal X}l(\boldsymbol x,\boldsymbol t)p(\boldsymbol x)\,d\boldsymbol x
=
\epsilon_{\mathrm F}^{(\boldsymbol t)}+\epsilon_{\mathrm{CF}}^{(\boldsymbol t)}.
\]
We integrate the conditional squared-loss identity over $p(\boldsymbol x)$ and apply
Equation~\eqref{app:cisi:marginal_risk_bound}.
The last inequality follows from $\nu_{\boldsymbol t}\geq0$.
\end{proof}

\begin{definition}
\label{def:cisi:coef_notation}
For each unordered pair of distinct treatment patterns $\{\boldsymbol u,\boldsymbol v\}\in\binom{\mathcal T}{2}$, define
\begin{align}
c_{\mathcal A}(\boldsymbol u,\boldsymbol v)
& \coloneqq 
\mathbb I[\boldsymbol u\in\mathcal A]p(\boldsymbol v)
+
\mathbb I[\boldsymbol v\in\mathcal A]p(\boldsymbol u),
\label{app:cisi:pair_weight}\\
\rho_{\mathcal A}
& \coloneqq 
\max_{\{\boldsymbol u,\boldsymbol v\}\in\binom{\mathcal T}{2}}
c_{\mathcal A}(\boldsymbol u,\boldsymbol v),
\label{app:cisi:rho}
\end{align}
where $\mathbb I[\cdot]$ is the indicator function.
\end{definition}

Because $\boldsymbol u$ and $\boldsymbol v$ are distinct treatment patterns, $0 \leq c_{\mathcal A}(\boldsymbol u,\boldsymbol v) \leq p(\boldsymbol u)+p(\boldsymbol v)\leq 1.$
Hence, $0\leq\rho_{\mathcal A}\leq1$.

\begin{thm}
\label{thm:cisi:average_contrast}
Suppose that Assumptions~\ref{ass:cisi:consistency}--\ref{ass:cisi:loss_class} hold for the squared loss and that the stated second moments are finite.
For any fixed coefficient vector in Definition~\ref{def:cisi:contrast},
\begin{equation}
\label{app:cisi:weighted_contrast_bound}
\mathcal E_{\boldsymbol a}
\leq\|\boldsymbol a\|_2^2
\left[
\sum_{\boldsymbol t\in\mathcal A}
\frac{\epsilon_{\mathrm F}^{(\boldsymbol t)}}{p(\boldsymbol t)}
+
B_{\Phi}
\sum_{\{\boldsymbol u,\boldsymbol v\}\in\binom{\mathcal T}{2}}
c_{\mathcal A}(\boldsymbol u,\boldsymbol v)D_{\boldsymbol u,\boldsymbol v}
\right].
\end{equation}
\end{thm}
\begin{proof}
For the predictor, the error in the average contrast is
\[
\hat\tau_{\boldsymbol a}-\tau_{\boldsymbol a}
=
\sum_{\boldsymbol t\in\mathcal A}a_{\boldsymbol t}
\mathbb E_{\boldsymbol X}[\delta_{\boldsymbol t}(\boldsymbol X)].
\]
The Cauchy--Schwarz inequality over the finite set $\mathcal A$ and Jensen's inequality over $\boldsymbol X$ yield
\begin{align*}
\mathcal E_{\boldsymbol a}
&\leq
\left(\sum_{\boldsymbol t\in\mathcal A}a_{\boldsymbol t}^2\right)
\sum_{\boldsymbol t\in\mathcal A}
\left(\mathbb E_{\boldsymbol X}[\delta_{\boldsymbol t}(\boldsymbol X)]\right)^2\\
&\leq\|\boldsymbol a\|_2^2
\sum_{\boldsymbol t\in\mathcal A}
\mathbb E_{\boldsymbol X}[\delta_{\boldsymbol t}(\boldsymbol X)^2]\\
&=\|\boldsymbol a\|_2^2\sum_{\boldsymbol t\in\mathcal A}\mathcal Q_{\boldsymbol t}\\
&\leq\|\boldsymbol a\|_2^2\sum_{\boldsymbol t\in\mathcal A}U_{\boldsymbol t},
\end{align*}
where the last step follows from Lemma~\ref{lem:cisi:squared_loss}.
The definition of $U_{\boldsymbol t}$ gives
\begin{equation*}
\mathcal E_{\boldsymbol a}
\leq\|\boldsymbol a\|_2^2
\left[
\sum_{\boldsymbol t\in\mathcal A}
\frac{\epsilon_{\mathrm F}^{(\boldsymbol t)}}{p(\boldsymbol t)}
+B_{\Phi}
\sum_{\boldsymbol t\in\mathcal A}
\sum_{\substack{\boldsymbol s\in\mathcal T\\\boldsymbol s\ne\boldsymbol t}}
p(\boldsymbol s)D_{\boldsymbol t,\boldsymbol s}
\right].
\end{equation*}
Since $D_{\boldsymbol t,\boldsymbol s}$ is symmetric, the double sum has the exact expression
\begin{equation*}
\sum_{\boldsymbol t\in\mathcal A}
\sum_{\boldsymbol s\ne\boldsymbol t}
p(\boldsymbol s)D_{\boldsymbol t,\boldsymbol s}
=
\sum_{\{\boldsymbol u,\boldsymbol v\}\in\binom{\mathcal T}{2}}
c_{\mathcal A}(\boldsymbol u,\boldsymbol v)D_{\boldsymbol u,\boldsymbol v}.
\end{equation*}
Indeed, an unordered pair contributes $p(\boldsymbol v)$ when $\boldsymbol u\in\mathcal A$, and it contributes $p(\boldsymbol u)$ when $\boldsymbol v\in\mathcal A$.
This proves Equation~\eqref{app:cisi:weighted_contrast_bound}.
\end{proof}

\paragraph{Connection to the pairwise balancing penalty.}
By Definitions~\ref{def:cisi:population_penalty} and \ref{def:cisi:coef_notation},
\begin{align*}
&\sum_{\{\boldsymbol u,\boldsymbol v\}\in\binom{\mathcal T}{2}}
c_{\mathcal A}(\boldsymbol u,\boldsymbol v)D_{\boldsymbol u,\boldsymbol v}\\
&\quad\leq\rho_{\mathcal A}
\sum_{\{\boldsymbol u,\boldsymbol v\}\in\binom{\mathcal T}{2}}
D_{\boldsymbol u,\boldsymbol v}
=\rho_{\mathcal A}C_{\mathcal T}\mathcal L_{\Phi}.
\end{align*}
Therefore, Theorem~\ref{thm:cisi:average_contrast} implies
\begin{equation}
\label{app:cisi:pair_average_contrast_bound}
\mathcal E_{\boldsymbol a}
\leq\|\boldsymbol a\|_2^2
\left[
\sum_{\boldsymbol t\in\mathcal A}
\frac{\epsilon_{\mathrm F}^{(\boldsymbol t)}}{p(\boldsymbol t)}
+B_{\Phi}\rho_{\mathcal A}C_{\mathcal T}\mathcal L_{\Phi}
\right].
\end{equation}
A weaker bound follows if $\rho_{\mathcal A}$ is replaced by one.
Thus, the average pairwise IPM controls the discrepancy term in the contrast bound.

\paragraph{A sharper bound with an outcome-noise correction.}
The proof also gives a sharper inequality if the term $\nu_{\boldsymbol t}$ is retained in Equation~\eqref{app:cisi:squared_loss_decomposition}:
\begin{equation*}
\mathcal E_{\boldsymbol a}
\leq\|\boldsymbol a\|_2^2
\left[
\sum_{\boldsymbol t\in\mathcal A}
\left\{\frac{\epsilon_{\mathrm F}^{(\boldsymbol t)}}{p(\boldsymbol t)}-\nu_{\boldsymbol t}\right\}
+B_{\Phi}
\sum_{\{\boldsymbol u,\boldsymbol v\}\in\binom{\mathcal T}{2}}
c_{\mathcal A}(\boldsymbol u,\boldsymbol v)D_{\boldsymbol u,\boldsymbol v}
\right].
\end{equation*}
Here, $\nu_{\boldsymbol t}$ is a conditional outcome-variance term averaged over covariates; it is not the sampling variance of the causal-effect estimator.
We omit this nonnegative noise correction from the stated corollaries so that their right-hand sides contain only factual prediction errors and representation discrepancies.

\subsection{ASE and AIE as Special Cases}
\label{app:cisi:ase_aie}

Let $\boldsymbol 0\in\mathcal T$ denote the all-zero treatment pattern, and let $\boldsymbol t_{+k}\in\mathcal T$ denote the pattern whose $k$-th component is one and whose other components are zero.
For $Q\subseteq\{1,\ldots,K\}$, let $\boldsymbol t_{(+Q)}$ have ones exactly at the indices in $Q$.
In particular, $\boldsymbol t_{(+\varnothing)}=\boldsymbol 0$.

\begin{cor}[Squared ASE error]
\label{cor:cisi:ase}
For any $k\in\{1,\ldots,K\}$, let $\mathcal E_{\mathrm{ASE}}(k)  \coloneqq \bigl(\hat\tau_{\mathrm{ASE}}(k)-\tau_{\mathrm{ASE}}(k)\bigr)^2$ denote the squared estimation error of ASE, and let $\mathcal A_k \coloneqq \{\boldsymbol t_{+k},\boldsymbol 0\}$.
Under the conditions of Theorem~\ref{thm:cisi:average_contrast},
\begin{equation*}
\mathcal E_{\mathrm{ASE}}(k)
\leq2\left[
\frac{\epsilon_{\mathrm F}^{(\boldsymbol t_{+k})}}{p(\boldsymbol t_{+k})}
+
\frac{\epsilon_{\mathrm F}^{(\boldsymbol 0)}}{p(\boldsymbol 0)}
+B_{\Phi}\rho_{\mathcal A_k}C_{\mathcal T}\mathcal L_{\Phi}
\right].
\end{equation*}
\end{cor}
\begin{proof}
The ASE has coefficients $a_{\boldsymbol t_{+k}}=1$ and $a_{\boldsymbol 0}=-1$, and all remaining coefficients are zero.
Hence, its support is $\mathcal A_k$ and $\|\boldsymbol a\|_2^2=2$.
Equation~\eqref{app:cisi:pair_average_contrast_bound} gives the result.
\end{proof}

\begin{cor}[Squared AIE error]
\label{cor:cisi:aie}
For any $S \subseteq \{1,\ldots,K\}$ with $|S|\geq 2$, let $E_{\mathrm{AIE}}(S) \coloneqq (\hat{\tau}_{\mathrm{AIE}}(S)-\tau_{\mathrm{AIE}}(S))^2$ denote the squared estimation error of AIE, and let $A_S  \coloneqq  \{\boldsymbol{t}_{(+Q)}:Q\subseteq S\}$.
Under the conditions of Theorem 1,
\begin{equation*}
\mathcal E_{\mathrm{AIE}}(S)
\leq2^{|S|}\left[
\sum_{Q\subseteq S}
\frac{\epsilon_{\mathrm F}^{(\boldsymbol t_{(+Q)})}}{p(\boldsymbol t_{(+Q)})}
+B_{\Phi}\rho_{\mathcal A_S}C_{\mathcal T}\mathcal L_{\Phi}
\right].
\end{equation*}

\end{cor}

\begin{proof}
Set
\[
a_{\boldsymbol t} \coloneqq 
\begin{cases}
(-1)^{|S|-|Q|},&\boldsymbol t=\boldsymbol t_{(+Q)}\text{ for }Q\subseteq S,\\
0,&\boldsymbol t\notin\mathcal A_S.
\end{cases}
\]
Distinct subsets $Q\subseteq S$ determine distinct treatment patterns.
Thus, $|\mathcal A_S|=2^{|S|}$, and every nonzero coefficient has absolute value one.
It follows that
\[
\|\boldsymbol a\|_2^2
=\sum_{Q\subseteq S}\left((-1)^{|S|-|Q|}\right)^2
=2^{|S|}.
\]
Equation~\eqref{app:cisi:pair_average_contrast_bound} now gives the result.
Only the patterns in $\mathcal A_S$ contribute factual errors to this bound.
\end{proof}

The bounds connect the squared errors of average single and interaction effects to treatment-conditional factual prediction errors and a pairwise IPM discrepancy.
The probability-weighted comparison in Equation~\eqref{app:cisi:weighted_contrast_bound} precedes the relaxation to the unweighted pair average.
This retains information about treatment-pattern frequencies and avoids a sum of factual errors over patterns that do not enter the contrast.
For example, under uniform pattern probabilities, $\rho_{\mathcal A}=2/|\mathcal T|$ whenever $|\mathcal A|\geq2$, so $\rho_{\mathcal A}C_{\mathcal T}=|\mathcal T|-1$.

By Definition~\ref{def:cisi:losses}, the ratio $\epsilon_{\mathrm F}^{(\boldsymbol t)}/p(\boldsymbol t)$ is a treatment-conditional expected squared loss; under consistency and ignorability, it equals the factual prediction errors conditional on $\boldsymbol T=\boldsymbol t$.
It is not, by itself, a lower bound on estimation difficulty or a variance formula for rare patterns.
The bounds motivate a combination of factual prediction loss and representation balancing, subject to the stated regularity conditions.

\section{Hardware and Software Environments}
\label{appendix:hard_soft_env}

We implement all methods in Python 3.12 and use TensorFlow 2.20.0 as the primary framework for the neural network models.
All experiments run on an x86\_64 system with Ubuntu 22.04.5 LTS.
The hardware configuration comprises 12 virtual CPUs backed by an Intel Xeon processor at 2.20\,GHz, 52\,GB of system memory, and a single NVIDIA L4 GPU with 23\,GB of GPU memory.

\section{Additional Simulation Results}
\label{appendix:additional_simulation_results}
\subsection{Additional Simulation Results under Homogeneous Treatment Effects}
This appendix reports experimental results using three simulation datasets with constant single and interaction treatment effects to evaluate the causal effect estimation performance of the proposed method.
The first scenario assumes that all true covariates are directly observable. 
The second scenario assumes that only proxy variables for latent covariates are observed. 
The third scenario represents a more realistic hybrid setting where true covariates and proxies for latent covariates coexist within the observed covariates, which combines the structures of the first and second scenarios. 
This allows us to test the generalizability of our method in complex settings.

In all scenarios, given the true covariates $\boldsymbol{x}_{i,\text{t}}$, treatment assignment and outcome generation follow the same functional form, with the number of treatments fixed at $K=3$ and the sample size set to $N=50,000$.
\begin{gather*}
t_{i}^{(k)} \sim \mathrm{Bern} \left( \sigma \left( \boldsymbol{w}_{t_k}^\top \boldsymbol{x}_{i,\text{t}} - \delta \right)  \right), \quad k=1,2,3,\\
\boldsymbol{t_i} = (t_{i}^{(1)},t_{i}^{(2)},t_{i}^{(3)}),\\
Y_i \sim \mathcal{N}\!\left(f\!\left(\boldsymbol{x}_{i,\mathrm{t}},\boldsymbol{t}_i\right),\,1^2\right),
\end{gather*}
where $\sigma(x) = 1 / (1 + \exp(-x))$, and $\delta$ is a bias parameter that varies across simulation datasets.
The vector $\boldsymbol{w}_{t_k}$ is the weight vector whose elements are independently drawn from the uniform distribution $U(-1,1)$. 
The outcome-generating function $f$ is defined as follows:
\begin{align*}
f(\boldsymbol{x}_{i,\text{t}}, \boldsymbol{t_i}) 
&= \boldsymbol{w}_x^\top \boldsymbol{x}_{i,\text{t}}
 + t_{i}^{(1)}
 + 1.2 t_{i}^{(2)} 
 + 0.8 t_{i}^{(3)}
 + t_{i}^{(1)} t_{i}^{(2)}
 -0.5 t_{i}^{(1)} t_{i}^{(3)}
 + 0.1 t_{i}^{(2)} t_{i}^{(3)}
 + 0.7 t_{i}^{(1)} t_{i}^{(2)}t_{i}^{(3)} + 2.
\end{align*}

We vary the structure of the observed covariates $\boldsymbol{x}_{i,\text{o}}$ and the true covariates $\boldsymbol{x}_{i,\text{t}}$ as follows:
\paragraph{Simulation dataset 1A}
\begin{gather*}
x_{i,n}^{(j)} \sim N(c_n^{(j)}, 1^2), \quad 
x_{i,u}^{(j)} \sim U(-10,10), \quad 
x_{i,b}^{(j)} \sim \mathrm{Bern}(p_b^{(j)}), \\
\boldsymbol{x}_{i,\text{o}} = \boldsymbol{x}_{i,\text{t}} = (\boldsymbol{x}_{i,n}, \boldsymbol{x}_{i,u}, \boldsymbol{x}_{i,b}),
\end{gather*}
where $j \in \{1,\dots,10\}$ and $c_n^{(j)}$ are drawn from the uniform distribution $U(-1,1)$. Similarly, each $p_b^{(j)}$ is sampled from $U(0,1)$. In simulation dataset 1A, $\delta = 0.3$.

\paragraph{Simulation dataset 2A}
\begin{gather*}
\boldsymbol{z}_i \sim N(\boldsymbol{0}, I_{10}), \quad
x_{i,n}^{(j)} \sim N(\boldsymbol{w}_n^{(j)\top} \boldsymbol{z}_i, 1^2), \\
x_{i,u}^{(j)} \sim N(\boldsymbol{w}_u^{(j)\top} \boldsymbol{z}_i, 5^2), \quad
x_{i,b}^{(j)} \sim \mathrm{Bern}(\sigma(\boldsymbol{w}_b^{(j)\top} \boldsymbol{z}_i)), \\
\boldsymbol{x}_{i,\text{o}} = (\boldsymbol{x}_{i,n}, \boldsymbol{x}_{i,u}, \boldsymbol{x}_{i,b}), \quad
\boldsymbol{x}_{i,\text{t}} = \boldsymbol{z}_i,
\end{gather*}
where $j \in \{1,\dots,10\}$ and the vectors $\boldsymbol{w}_n^{(j)}$, $\boldsymbol{w}_u^{(j)}$, and $\boldsymbol{w}_b^{(j)}$ are weight vectors whose elements are drawn independently from $U(-1,1)$. $I_d$ denotes the identity matrix $d \times d$, which is also used as the variance-covariance matrix. In simulation dataset 2A, $\delta = 0.2$.

\paragraph{Simulation dataset 3A}
\begin{gather*}
\boldsymbol{z}_i \sim N(\boldsymbol{0}, I_{15}),\quad 
x_{i,p}^{(l)} \sim N(\boldsymbol{w}_p^{(l)\top} \boldsymbol{z}_i, 1^2),\\
x_{i,n}^{(j)} \sim N(c_n^{(j)}, 1^2),\quad 
x_{i,u}^{(j)} \sim U(-10,10),\quad 
x_{i,b}^{(j)} \sim \mathrm{Bern}(p_b^{(j)}), \\
\boldsymbol{x}_{i,\text{o}} = (\boldsymbol{x}_{i,n}, \boldsymbol{x}_{i,u}, \boldsymbol{x}_{i,b}, \boldsymbol{x}_{i,p}), \quad
\boldsymbol{x}_{i,\text{t}} = (\boldsymbol{x}_{i,n}, \boldsymbol{x}_{i,u}, \boldsymbol{x}_{i,b}, \boldsymbol{z}_i),
\end{gather*}
where $l \in \{1,\dots,15\}$ and $j \in \{1,\dots,5\}$. Vectors $\boldsymbol{w}_p^{(l)}$ are weight vectors whose elements are drawn independently from $U(-1,1)$. In simulation dataset 3A, $\delta = 0.1$.

Table~\ref{tab:simulation_result_linear_all} shows the $\epsilon_{\mathrm{ASE}}$ and $\epsilon_{\mathrm{AIE}}$ of the proposed and baseline methods in three simulation datasets. 
In all datasets, the proposed method achieves the lowest estimation errors.
These results demonstrate CISI-Net's ability to accurately estimate single treatment and interaction treatment effects even when no heterogeneity exists in their effects.

\begin{table}[ht]
\caption{Comparison of mean and standard deviation of $\epsilon_{\mathrm{ASE}}$ and $\epsilon_{\mathrm{AIE}}$ across simulation datasets. Here, $k \in \{1,2,3\}$ indexes single treatments for ASE, and $S \subseteq \{1,2,3\}$ (with $|S| \ge 2$) denotes treatment combinations for AIE.}
\label{tab:simulation_result_linear_all}
\centering
\fontsize{19pt}{22pt}\selectfont
\setlength{\tabcolsep}{13pt} 
\renewcommand{\arraystretch}{1.5}

\resizebox{\textwidth}{!}{       
\begin{tabular}{p{0.4cm}lccc|cccc}
\hline
\multicolumn{2}{l}{} & \multicolumn{3}{c|}{$\epsilon_{\mathrm{ASE}}$} & \multicolumn{4}{c}{$\epsilon_{\mathrm{AIE}}$} \\
\cline{3-5}\cline{6-9}
Sim.&Method& $k=1$ & $k=2$ & $k=3$ & $S=\{1,2\}$ & $S=\{2,3\}$ & $S=\{1,3\}$ & $S=\{1,2,3\}$ \\
\hline
\multirow{5}{*}{\centering 1A}
& TARNet       & 0.19 $\pm$ 0.12  & 0.17 $\pm$ 0.13  & 0.16 $\pm$ 0.13  & 0.27 $\pm$ 0.18  & 0.25 $\pm$ 0.19  & 0.24 $\pm$ 0.17  & 0.41 $\pm$ 0.31 \\
& CFR-WASS     & 0.19 $\pm$ 0.13  & 0.17 $\pm$ 0.13  & 0.17 $\pm$ 0.13  & 0.28 $\pm$ 0.23  & 0.26 $\pm$ 0.23  & 0.26 $\pm$ 0.19  & 0.47 $\pm$ 0.38 \\
& NCoRE        & 0.10 $\pm$ 0.08& 0.09 $\pm$ 0.10& 0.09 $\pm$ 0.08& 0.15 $\pm$ 0.13  & 0.14 $\pm$ 0.11  & 0.16 $\pm$ 0.14  & 0.20 $\pm$ 0.15 \\
& TECE-VAE     & 0.43 $\pm$ 0.10 & 0.47 $\pm$ 0.10  & 0.37 $\pm$ 0.07 & 0.56 $\pm$ 0.20  & 0.08 $\pm$ 0.07  & 0.41 $\pm$ 0.11  & 0.73 $\pm$ 0.10 \\
& CISI-Net & \fontseries{b}\selectfont{0.04 $\pm$ 0.03} & \fontseries{b}\selectfont{0.04 $\pm$ 0.03} & \fontseries{b}\selectfont{0.04 $\pm$ 0.03} & \fontseries{b}\selectfont{0.05 $\pm$ 0.04} & \fontseries{b}\selectfont{0.04 $\pm$ 0.03} & \fontseries{b}\selectfont{0.05 $\pm$ 0.03} & \fontseries{b}\selectfont{0.06 $\pm$ 0.04} \\
\hline
\multirow{5}{*}{\centering 2A}
& TARNet       & 0.05 $\pm$ 0.04 & 0.06 $\pm$ 0.04 & 0.05 $\pm$ 0.04 & 0.09 $\pm$ 0.07 & 0.08 $\pm$ 0.07 & 0.08 $\pm$ 0.06 & 0.14 $\pm$ 0.11 \\
& CFR-WASS     & 0.07 $\pm$ 0.06 & 0.07 $\pm$ 0.06 & 0.07 $\pm$ 0.05 & 0.11 $\pm$ 0.09 & 0.10 $\pm$ 0.07 & 0.16 $\pm$ 0.07 & 0.17 $\pm$ 0.13 \\
& NCoRE        & 0.05 $\pm$ 0.04 & 0.05 $\pm$ 0.04 & 0.04 $\pm$ 0.03 & 0.09 $\pm$ 0.07 & 0.08 $\pm$ 0.06 & 0.08 $\pm$ 0.06 & 0.12 $\pm$ 0.11 \\
& TECE-VAE     & 0.09 $\pm$ 0.06 & 0.11 $\pm$ 0.07 & 0.09 $\pm$ 0.06 & 0.12 $\pm$ 0.05 & 0.11 $\pm$ 0.04 & 0.09 $\pm$ 0.04 & 0.13 $\pm$ 0.09 \\
& CISI-Net & \fontseries{b}\selectfont{0.04 $\pm$ 0.03} & \fontseries{b}\selectfont{0.05 $\pm$ 0.03} & \textbf{0.04 $\pm$ 0.03} & \fontseries{b}\selectfont{0.04 $\pm$ 0.03} & \fontseries{b}\selectfont{0.03 $\pm$ 0.02} & \fontseries{b}\selectfont{0.03 $\pm$ 0.02} & \fontseries{b}\selectfont{0.04 $\pm$ 0.03} \\
\hline
\multirow{5}{*}{\centering 3A}
& TARNet & 0.35 $\pm$ 0.26 & 0.31 $\pm$ 0.25 & 0.32 $\pm$ 0.26 & 0.24 $\pm$ 0.19 & 0.24 $\pm$ 0.19 & 0.24 $\pm$ 0.19 & 0.36 $\pm$ 0.33 \\
& CFR-WASS & 0.29 $\pm$ 0.23 & 0.27 $\pm$ 0.22 & 0.29 $\pm$ 0.22 & 0.23 $\pm$ 0.19 & 0.24 $\pm$ 0.19 & 0.24 $\pm$ 0.18 & 0.40 $\pm$ 0.32 \\
& NCoRE & 0.38 $\pm$ 0.29 & 0.35 $\pm$ 0.26 & 0.30 $\pm$ 0.26 & 0.28 $\pm$ 0.23 & 0.24 $\pm$ 0.18 & 0.23 $\pm$ 0.18 & 0.36 $\pm$ 0.33 \\
& TECE-VAE & 0.36 $\pm$ 0.18 & 0.43 $\pm$ 0.18 & 0.31 $\pm$ 0.19 & 0.17 $\pm$ 0.13 & 0.13 $\pm$ 0.09 & 0.29 $\pm$ 0.11 & 0.58 $\pm$ 0.15 \\
& CISI-Net & \fontseries{b}\selectfont{0.18 $\pm$ 0.14} & \fontseries{b}\selectfont{0.17 $\pm$ 0.13} & \fontseries{b}\selectfont{0.18 $\pm$ 0.15} & \fontseries{b}\selectfont{0.11 $\pm$ 0.09} & \fontseries{b}\selectfont{0.08 $\pm$ 0.06} & \fontseries{b}\selectfont{0.08 $\pm$ 0.07} & \fontseries{b}\selectfont{0.11 $\pm$ 0.10} \\
\hline
\end{tabular}}
\end{table}

\subsection{Scalability to a Larger Number of Treatments}
\label{appendix:scalability_treatment_num_exp}
We investigate whether the proposed method maintains its performance advantage over existing methods designed for multiple treatment settings as the number of treatments increases.
With $K$ treatments, the total number of possible treatment patterns grows exponentially as $2^{K}$, and the number of possible interaction effects increases accordingly.
Scenarios requiring the estimation of a large number of causal effects frequently arise in real-world applications~\citep{parmar2008speeding, freidlin2008multi}.
Therefore, we evaluate the practical utility of each method by assessing its accuracy of estimation across varying numbers of treatments.

To examine the scalability of each method, we conduct experiments under varying numbers of treatments $K \in \{4,5,8,10\}$.
In practice, the number of simultaneous treatments rarely exceeds ten~\citep{cho2022aging, freidlin2008multi, muralidharan2025factorial, Tsuboi31122024}. 
Therefore, our simulation setting with an upper bound of $K \le 10 $ represents a practical range that adequately covers the vast majority of multiple-treatment scenarios.
For all settings, treatment assignment and outcome generation follow the same functional form, and the sample size is fixed at $N=50,000$.
The treatment vector, covariates, and outcomes are based on the following data generation process:
\begin{gather*}
x_{i,n}^{(j)} \sim N(c_n^{(j)}, 1^2), \quad 
x_{i,u}^{(j)} \sim U(-1,1), \quad 
\boldsymbol{x}_{i} = (\boldsymbol{x}_{i,n}, \boldsymbol{x}_{i,u}),\\
H_{i}^{(v)} = \mathbb{I}\left(\,x_{i}^{(v)} + x_{i}^{(v+1)} > 1\,\right), \quad 
\boldsymbol{H}_{i} = (H_{i}^{(1)}, H_{i}^{(2)}),\\
t_{i}^{(k)} \sim \mathrm{Bern} \left( \sigma \left( \boldsymbol{w}_{t_k}^\top \boldsymbol{x}_{i}- \boldsymbol{w}_{H}^{\top}\boldsymbol{H}_{i} \right) \right),\quad
\boldsymbol{t}_{i} = (t_{i}^{(1)},\dots,t_{i}^{(K)}),\\
Y_i \sim \mathcal{N}\!\left(g\!\left(\boldsymbol{x}_{i},\boldsymbol{t}_i\right),\,1^2\right),
\end{gather*}
where $j \in \{1,\dots,15\}$ denotes the covariate index, $v \in \{1,2\}$, and $k \in \{1,\dots,K\}$.
$\mathbb{I}(\cdot)$ is the indicator function, $\sigma(\cdot)$ is the sigmoid function defined as $\sigma(x) = 1 / (1 + \exp(-x))$.
The vectors $\boldsymbol{w}_{t_k}$ and $\boldsymbol{w}_{H}$ are the weight vectors whose elements are independently drawn from the uniform distribution $U(-1,1)$.
We set $\boldsymbol{x}_{i,\text{o}} = \boldsymbol{x}_{i,\text{t}} = (\boldsymbol{x}_{i,n}, \boldsymbol{x}_{i,u})$.
The outcome generating function $g$ is defined as follows:
\begin{gather*}
g(\boldsymbol{x}_{i},\boldsymbol{t}_i)
=
\boldsymbol{w}_{x}^{\top}\boldsymbol{x}_{i}
+
\sum_{\substack{\emptyset \neq S \subseteq \{1,\dots,K\} \\ |S| \le K}}
\tau_S(\boldsymbol{x}_i)\prod_{k\in S} t_{i}^{(k)}
+2, \ \ 
\tau_S(\boldsymbol{x}_i) = w_{S}  \left(1 + x_i^{(j_{S})}\right),
\end{gather*}
where $j_{S} \in \{1,\dots,30\}$ denotes a covariate index randomly assigned to each subset $S$.
The vector $\boldsymbol{w}_{x}$ is a weight vector whose elements are independently drawn from the uniform distribution $U(-1,1)$, and each $w_{S}$ is independently drawn from $U(-2,2)$.
This setup induces treatment assignment bias through $\boldsymbol{H}_i$ and heterogeneous treatment effects through $\tau_S(\boldsymbol{x}_i)$.

The hyperparameters specified in Section~\ref{subsec:implement_details} are applied to CISI-Net and the multiple-treatment baseline methods NCoRE and TECE-VAE.
Table~\ref{tab:parameter_count_varying_K} shows the number of trainable parameters for each method as a reference for model capacity.

\begin{table}[t]
\centering
\caption{Comparison of the number of trainable parameters across methods under different numbers of treatments $K$. M denotes $10^6$.}
\label{tab:parameter_count_varying_K}
\renewcommand{\arraystretch}{1.2}
\begin{tabular}{lccccc}
\hline
$K$ & TECE-VAE & NCoRE & CISI-Net \\
\hline
4  & 0.425M & 0.518M  & 0.137M \\
5  & 0.426M & 0.842M  & 0.137M \\
8  & 0.429M & 2.298M & 0.138M \\
10 & 0.430M & 3.670M  & 0.138M \\
\hline
\end{tabular}
\end{table}

\begin{figure}[tb]
  \centering
  \includegraphics[width=\linewidth, keepaspectratio]{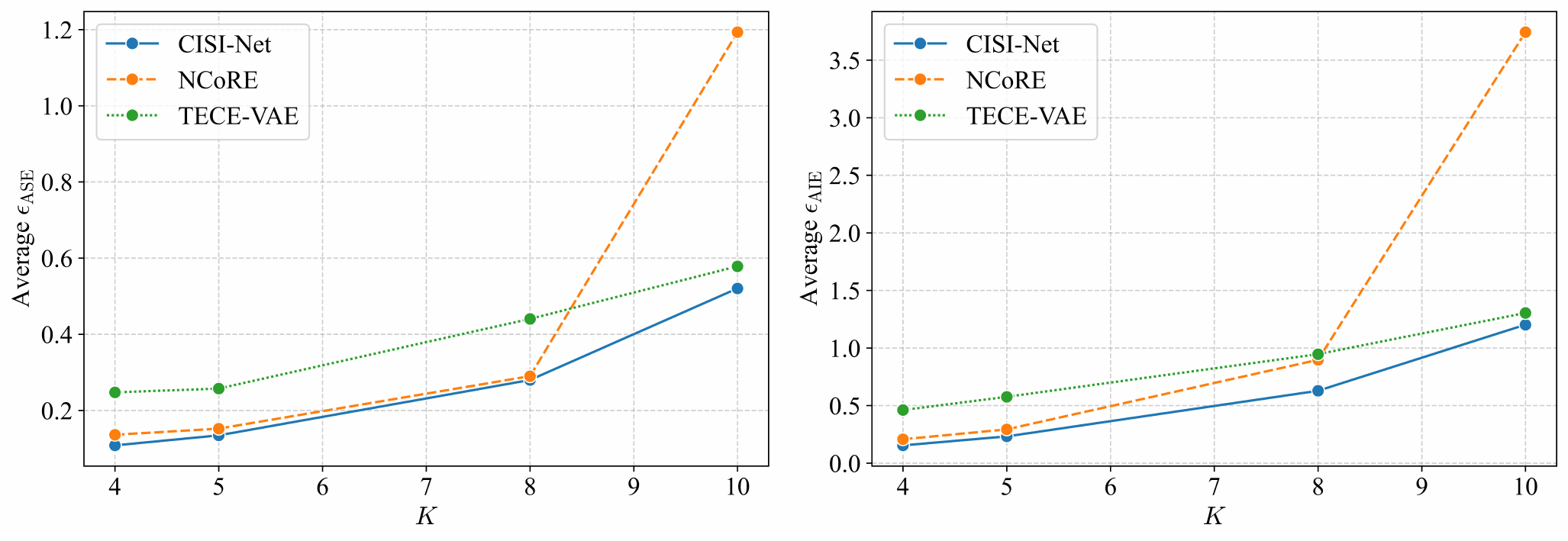}
  \caption{Average $\epsilon_{\mathrm{ASE}}$~(left) and $\epsilon_{\mathrm{AIE}}$~(right) under varying numbers of treatments $K \in \{4,5,8,10\}$.}
  \label{fig:ase_aie_errors_treatmentNum}
\end{figure}

Figure~\ref{fig:ase_aie_errors_treatmentNum} shows the average $\epsilon_{\mathrm{ASE}}$ and $\epsilon_{\mathrm{AIE}}$ across varying numbers of treatments $K$. Estimation errors are averaged over all single effects and interaction effects, respectively.
CISI-Net consistently achieves lower estimation errors than the baseline methods for both ASE and AIE, and maintains its superior performance as $K$ increases.
These results suggest that CISI-Net provides a stable and accurate estimate as the number of treatments increases, demonstrating its practical utility.

The baseline methods, NCoRE and TECE-VAE, underperform CISI-Net, suggesting limitations inherent in their respective architectures. 
Whereas NCoRE demonstrates competitive performance on both metrics for $K \leq 8$, its performance deteriorates markedly at $K = 10$. 
This behavior reflects a limitation of the NCoRE architecture, which constructs outcome and interaction networks independently and does not adequately promote parameter sharing across treatment patterns.
As the number of treatment patterns increases, insufficient parameter sharing can result in an unstable estimation.
Although TECE-VAE shows a more gradual increase in both $\epsilon_{\mathrm{ASE}}$ and $\epsilon_{\mathrm{AIE}}$, its estimation error remains consistently higher than that of CISI-Net, especially for $K \le 8$.
Unlike NCoRE, TECE-VAE employs task embeddings and does not construct independent networks for each treatment pattern. 
As a result, performance degradation is mitigated as $K$ increases.
Despite this advantage, reliance on latent-variable modeling under potential model misspecification may lead to unnecessary latent recovery, which can result in inferior estimation accuracy compared to CISI-Net.

\subsection{Comparison of Computing Cost}
\label{appendix:computing_cost}
In this section, we investigate a computational limitation of CISI-Net arising from the increasing number of pairwise discrepancy evaluations as the number of treatment patterns grows.
The pairwise balancing penalty in CISI-Net requires evaluating discrepancies between treatment-pattern-specific representation distributions.
If the number of treatments is $K$, the number of possible treatment patterns is $L=2^K$, and the number of unordered treatment-pattern pairs is $\binom{L}{2}=O(L^2)$.
Therefore, although CISI-Net does not construct a separate outcome network for each treatment pattern, the computational cost of the pairwise balancing penalty can increase substantially as the number of treatments grows.

To empirically evaluate this computational cost, we measure the wall-clock training time per epoch under the same simulation setting used in Appendix~\ref{appendix:scalability_treatment_num_exp}.
We vary the number of treatments as $K \in \{3,4,5,8,10\}$ while keeping the remaining training configuration fixed.
All experiments are conducted under the hardware and software environment described in Appendix~\ref{appendix:hard_soft_env}.
For each value of $K$, we conduct 30 trials and report the training time per epoch averaged across all observed epochs.

\begin{table}[tb]
  \centering
  \caption{
  Wall-clock training time per epoch~(s/epoch) for CISI-Net under varying numbers of treatments $K$.
  Times are averaged across all observed epochs in 30 trials under the same hardware and training configuration.}
  \renewcommand{\arraystretch}{1.1}
  \begin{tabular}{ccccc}
    \toprule
    $K=3$ & $K=4$ & $K=5$ & $K=8$ & $K=10$ \\
    \midrule
    14.90 & 33.12 & 136.44 & 1093.98 & 2229.83 \\
    \bottomrule
  \end{tabular}
  \label{tab:compute_cost}
\end{table}

Table~\ref{tab:compute_cost} shows that the training time increases substantially as the number of treatments grows.
The training time increases from 14.90~s/epoch at $K=3$ to 136.44~s/epoch at $K=5$, and further increases to 1093.98~s/epoch at $K=8$ and 2229.83~s/epoch at $K=10$.
This increase is consistent with the growth in the number of treatment-pattern pairs involved in the pairwise balancing computation.
Thus, although the results in Appendix~\ref{appendix:scalability_treatment_num_exp} show that CISI-Net maintains its estimation performance as the number of treatments increases, the computational cost of pairwise balancing remains a limitation when the number of treatment patterns becomes large.

\subsection{Additional Ablation Results}
\label{appendix:additional_ablation_results}
We conduct additional ablation studies using simulation datasets that differ from the one used in Section~\ref{sec:simulation_results} to verify the dataset-independent utility of the balancing penalty in CISI-Net.
Specifically, we use simulation datasets 2 and 3, which are defined in Section~\ref{subsec:simulation_datasets_gdp}. 
To ensure robustness against randomness in data generation, we generate 100 independent datasets for each scenario using different random seeds and report the average $\epsilon_{\mathrm{ASE}}$ and $\epsilon_{\mathrm{AIE}}$ across these datasets for each model.
Table~\ref{tab:additional_simulation_ablation} shows the results of the ablation study for each dataset.

Overall, the results demonstrate that applying the balancing penalty improves the accuracy of causal effect estimation across the two additional simulation datasets.
In simulation dataset 2, the improvement from the balancing penalty is relatively small.
This may be because the observed covariates serve as proxies for latent covariates, which partially limits the benefits of balancing representation distributions due to the incomplete information contained in these proxies.
Nevertheless, CISI-Net with the balancing penalty achieves equal or lower errors compared to CISI-Net without the balancing penalty across most metrics.
These results suggest that the contribution of the balancing penalty is not limited to a specific dataset but supports stable causal effect estimation across different data-generating processes, regardless of the presence of interaction effects.

\begin{table*}[tb]
\caption{Additional results comparing CISI-Net with and without the balancing penalty~(BP) on simulation datasets 2 and 3. Here, $k \in \{1,2,3\}$ indexes single treatments for ASE, and $S \subseteq \{1,2,3\}$ with $|S| \ge 2$ denotes treatment combinations for AIE. The balancing coefficient $\alpha$ is set to 0.1 when the balancing penalty is applied.}
\label{tab:additional_simulation_ablation}
\centering
\setlength{\tabcolsep}{5pt}
\renewcommand{\arraystretch}{1.35}
\resizebox{\textwidth}{!}{
\begin{tabular}{llccc|cccc}
\hline
\multicolumn{2}{l}{} 
& \multicolumn{3}{c|}{$\epsilon_{\mathrm{ASE}}$} 
& \multicolumn{4}{c}{$\epsilon_{\mathrm{AIE}}$} \\
\cline{3-5}\cline{6-9}
Sim. & Method
& $k=1$ & $k=2$ & $k=3$ 
& $S=\{1,2\}$ & $S=\{2,3\}$ & $S=\{1,3\}$ & $S=\{1,2,3\}$ \\
\hline

\multirow{2}{*}{2}
& CISI-Net w/o BP
& 0.16 $\pm$ 0.13 
& 0.20 $\pm$ 0.13 
& 0.12 $\pm$ 0.10 
& 0.27 $\pm$ 0.15 
& 0.23 $\pm$ 0.14 
& 0.19 $\pm$ 0.12 
& 0.21 $\pm$ 0.13 \\

& CISI-Net
& 0.16 $\pm$ 0.13
& 0.19 $\pm$ 0.13
& 0.12 $\pm$ 0.09
& 0.26 $\pm$ 0.18
& 0.23 $\pm$ 0.14
& 0.18 $\pm$ 0.12
& 0.21 $\pm$ 0.14 \\
\hline

\multirow{2}{*}{3}
& CISI-Net w/o BP
& 0.11 $\pm$ 0.07 
& 0.11 $\pm$ 0.08 
& 0.09 $\pm$ 0.07 
& 0.18 $\pm$ 0.15 
& 0.16 $\pm$ 0.11 
& 0.17 $\pm$ 0.11 
& 0.09 $\pm$ 0.06 \\

& CISI-Net
& 0.10 $\pm$ 0.06
& 0.09 $\pm$ 0.08
& 0.08 $\pm$ 0.06
& 0.12 $\pm$ 0.09
& 0.12 $\pm$ 0.08
& 0.10 $\pm$ 0.08
& 0.09 $\pm$ 0.07 \\
\hline
\end{tabular}
}
\end{table*}

\section{Details and Diagnostics of Real-World Datasets}
\label{appendix:real_world_diagnostics}
This section provides additional diagnostics for the two real-world marketing datasets described in Section~\ref{sec:real_world_application} to characterize the empirical support relevant to causal effect estimation.
We examine three complementary aspects because reliable causal effect estimation from observational data requires sufficient empirical support for each treatment pattern across covariate strata and adequate comparability of observed covariate distributions across treatment patterns~\citep{imai2004causal,austin2009balance,petersen2012diagnosing}.
First, we report the observed frequencies of all treatment patterns to characterize treatment-pattern sparsity.
Second, we assess covariate balance across treatment patterns using pairwise absolute standardized mean differences~(SMDs), which are widely used to diagnose differences in observed baseline covariates across treatment groups~\citep{austin2009balance}.
Third, we examine practical overlap using estimated generalized propensity scores~\citep{imai2004causal}.

\paragraph{Treatment-pattern frequencies.}
Table~\ref{tab:treatment_pattern_proportions} reports the observed proportions of treatment patterns in the two datasets.
Dataset 1 consists of eight treatment patterns corresponding to three marketing promotions, with substantial variation in their observed frequencies.
Among the patterns involving multiple simultaneous promotions, $(1,1,0)$ and $(1,1,1)$ were particularly rare, accounting for only 0.15\% and 0.19\% of the observations, respectively.
Dataset 2 consists of four treatment patterns, with $(1,1)$ representing simultaneous assignment of both promotions, accounting for only 0.94\% of the observations.
These results indicate that treatment-pattern sparsity is pronounced for some simultaneous multiple-promotion assignments.

\begin{table}[tb]
\centering
\caption{
Observed proportions of treatment patterns in the two real-world marketing datasets.
}
\begin{tabular}{cccc@{\hspace{1.2cm}}cc}
\toprule
\multicolumn{4}{c}{Dataset 1}
&
\multicolumn{2}{c}{Dataset 2}
\\
\cmidrule(lr){1-4}
\cmidrule(lr){5-6}

Treatment pattern & Share (\%)
& Treatment pattern & Share (\%)
& Treatment pattern & Share (\%)
\\
\midrule

$(0,0,0)$ & 36.52
& $(1,1,0)$ & 0.15
& $(0,0)$ & 64.70
\\

$(1,0,0)$ & 2.00
& $(0,1,1)$ & 2.76
& $(1,0)$ & 29.08
\\

$(0,1,0)$ & 15.02
& $(1,0,1)$ & 1.25
& $(0,1)$ & 5.28
\\

$(0,0,1)$ & 42.11
& $(1,1,1)$ & 0.19
& $(1,1)$ & 0.94
\\

\bottomrule
\end{tabular}
\label{tab:treatment_pattern_proportions}
\end{table}
\paragraph{Covariate-balance diagnostics.}
Table~\ref{tab:covariate_balance_summary} summarizes pairwise covariate-balance diagnostics by reporting the absolute $|\mathrm{SMD}|$ for each of the 71 observed covariates across all unique pairs of treatment patterns.
In Dataset 1, the median $|\mathrm{SMD}|$ across all treatment-pattern pairs and covariates was 0.061, while 35.5\% and 17.1\% of the comparisons exceeded 0.1 and 0.2, respectively.
The largest pairwise imbalance was observed between $(0,0,0)$ and $(1,1,1)$, for which the median $|\mathrm{SMD}|$ was 0.128 and 59.6\% of the covariates had $|\mathrm{SMD}|>0.1$.
In Dataset 2, the overall median $|\mathrm{SMD}|$ was 0.086, and 46.8\% of the comparisons exceeded 0.1.
The largest median $|\mathrm{SMD}|$ in Dataset 2 was observed for $(0,0)$ versus $(1,1)$, with a value of 0.149.
Thus, the observed covariate distributions are not perfectly balanced across treatment patterns, and the degree of imbalance differs across pattern pairs.

\begin{table}[t]
\centering
\caption{
Summary of pairwise covariate-balance diagnostics across treatment patterns.
Pairwise $|\mathrm{SMD}|$ values are computed across all 71 observed covariates for each unique treatment-pattern pair.
}
\begin{tabular}{lccccc}
\toprule
Dataset
& Median $|\mathrm{SMD}|$
& 90th pct. $|\mathrm{SMD}|$
& $\%$ with $|\mathrm{SMD}|>0.1$
& $\%$ with $|\mathrm{SMD}|>0.2$
& Max. $|\mathrm{SMD}|$ \\
\midrule
Dataset 1 & 0.061 & 0.270 & 35.5 & 17.1 & 1.632 \\
Dataset 2 & 0.086 & 0.339 & 46.8 & 25.3 & 1.760 \\
\bottomrule
\end{tabular}
\label{tab:covariate_balance_summary}
\end{table}

\begin{figure}[tb]
  \centering
  \begin{subfigure}[b]{\linewidth}
    \centering
    \includegraphics[width=\linewidth, keepaspectratio]{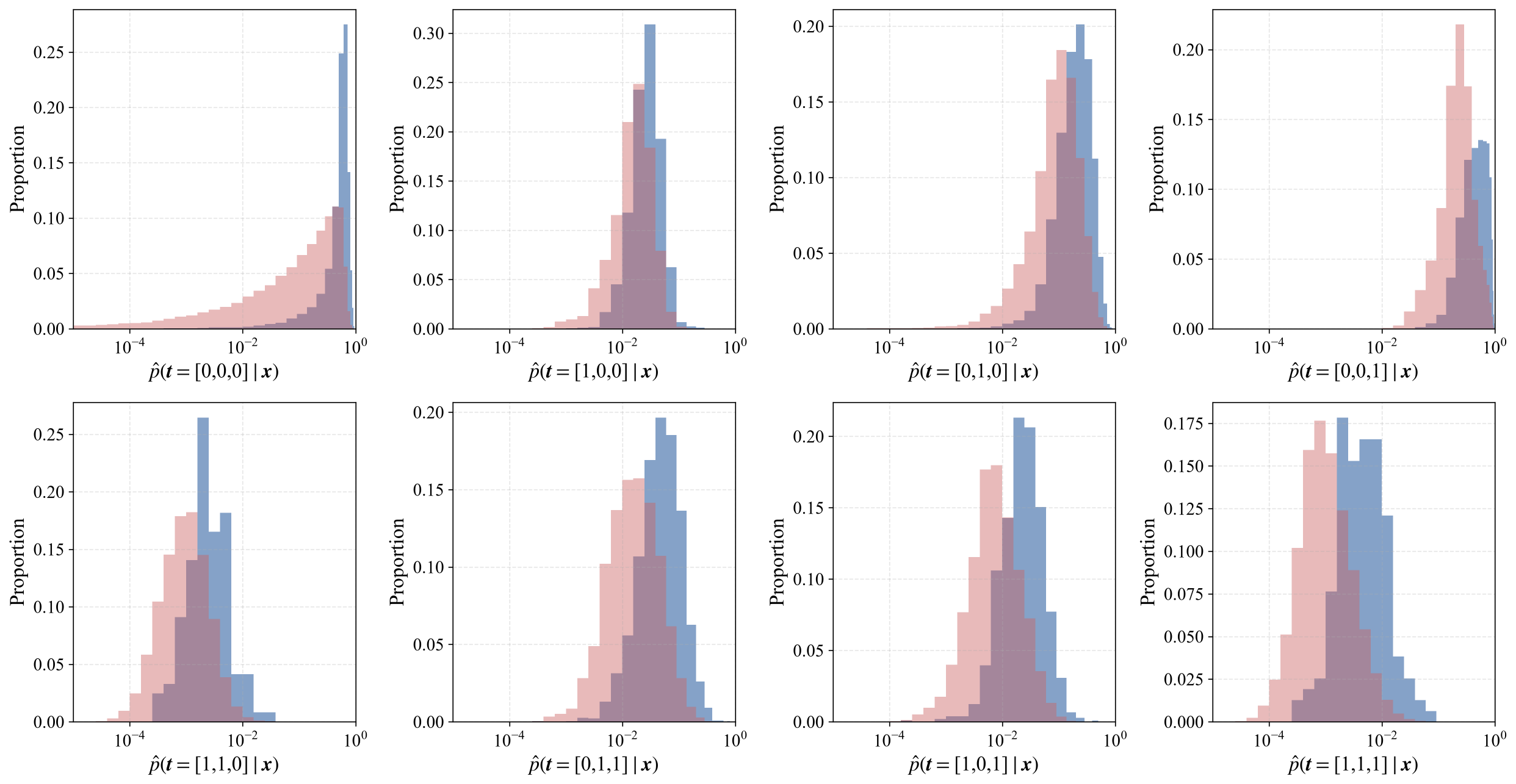}
    \caption{Dataset 1.}
    \label{fig:overlap_dataset1}
  \end{subfigure}

  \vskip\baselineskip

  \begin{subfigure}[b]{\linewidth}
    \centering
    \includegraphics[width=\linewidth, keepaspectratio]{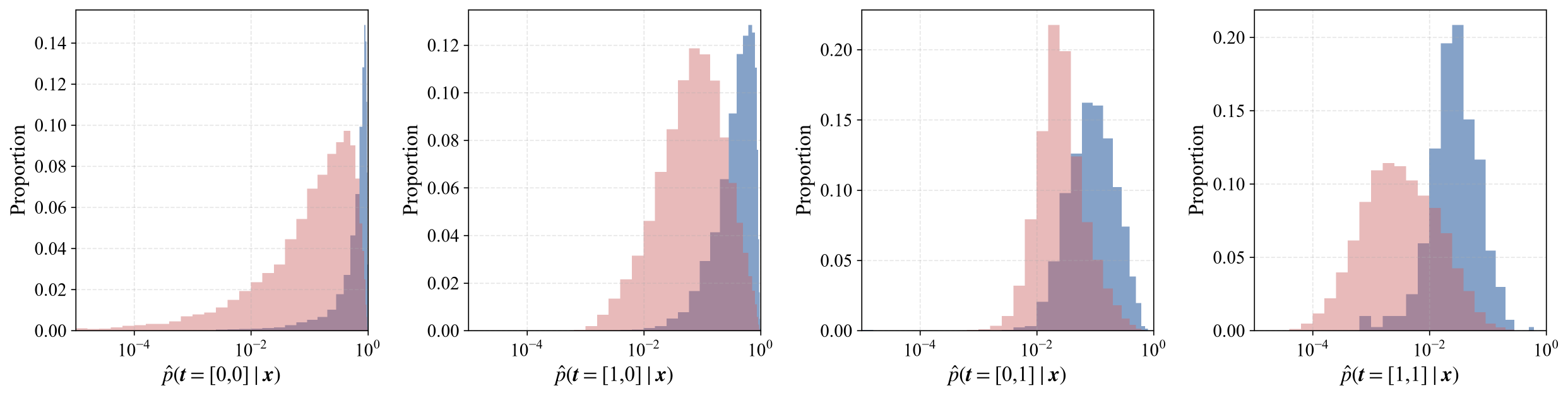}
    \caption{Dataset 2.}
    \label{fig:overlap_dataset2}
  \end{subfigure}

  \caption{
    Practical-overlap diagnostics based on estimated generalized propensity scores for the two real-world marketing datasets.
    For each treatment pattern $\boldsymbol{t}$, the distributions of $\hat{p}(\boldsymbol{t}\mid\boldsymbol{x})$ are compared between units who received treatment pattern $\boldsymbol{t}$~(blue; $\boldsymbol{T}_i=\boldsymbol{t}$) and those who received other treatment patterns~(red; $\boldsymbol{T}_i\neq\boldsymbol{t}$).
    The horizontal axis is shown on a logarithmic scale.
}
  \label{fig:practical_overlap_diagnostics}
\end{figure}

\paragraph{Practical-overlap diagnostics.}
Finally, we assessed practical overlap using the estimated generalized propensity score $\hat{p}(\boldsymbol{t}\mid\boldsymbol{x})$, which denotes the probability of receiving treatment pattern $\boldsymbol{t}$ conditional on the observed covariates $\boldsymbol{x}$~\citep{imai2004causal}.
Figure~\ref{fig:practical_overlap_diagnostics} shows the distributions of $\hat{p}(\boldsymbol{t}\mid\boldsymbol{x})$ for units who received treatment pattern $\boldsymbol{t}$~(blue; $\boldsymbol{T}_i=\boldsymbol{t}$) and those who received other treatment patterns~(red; $\boldsymbol{T}_i\neq\boldsymbol{t}$).

Across most treatment patterns, the two distributions exhibit overlapping support, although their locations and the extent of overlap differ across patterns.
The differences are more pronounced for some treatment patterns, while no pattern exhibits clear complete separation based on the observed covariates.
Together with the substantial sparsity of several multiple-treatment patterns shown in Table~\ref{tab:treatment_pattern_proportions}, these results indicate that practical overlap varies across treatment patterns and should be considered when interpreting the estimated causal effects.

\section{Sensitivity Analysis in Empirical Applications}
\label{appendix:sensitivity_analysis_in_empirical_applications}
\begin{figure}[t!]
  \centering

  \begin{subfigure}[t]{\linewidth}
    \centering
    \includegraphics[width=\linewidth, keepaspectratio]{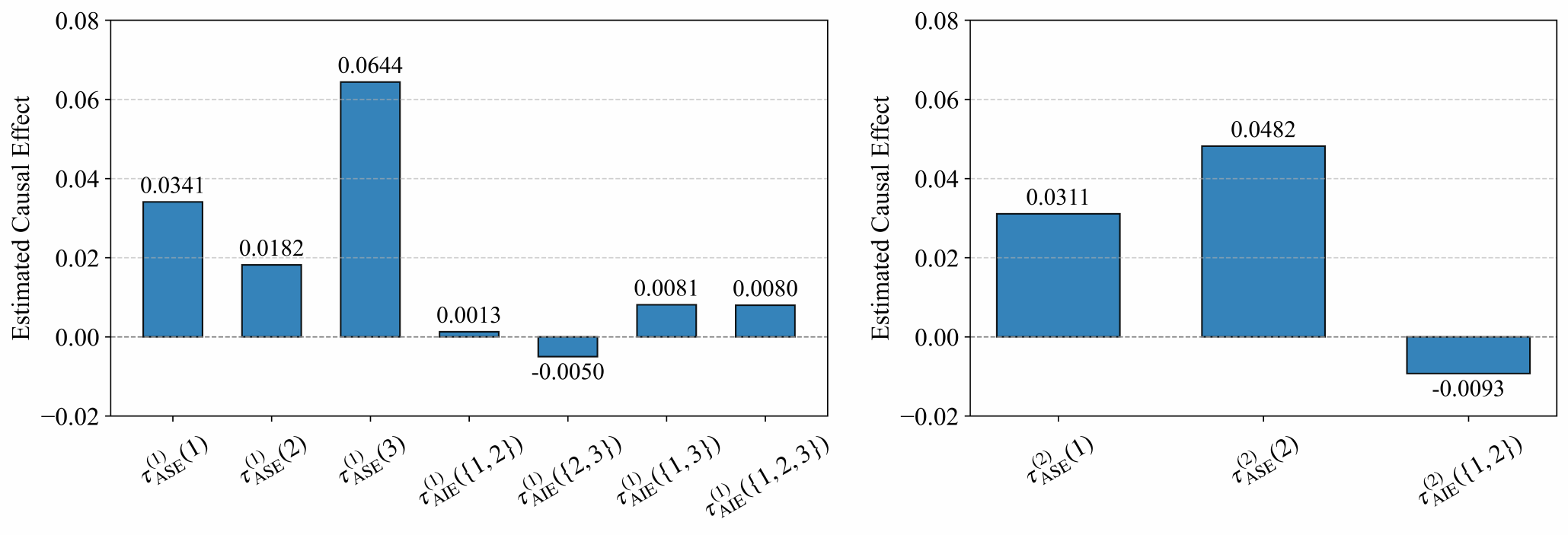}
    \caption{$\alpha=0.5$}
    \label{fig:empirical_results_alpha_05}
  \end{subfigure}

  \vspace{0.6em}

  \begin{subfigure}[t]{\linewidth}
    \centering
    \includegraphics[width=\linewidth, keepaspectratio]{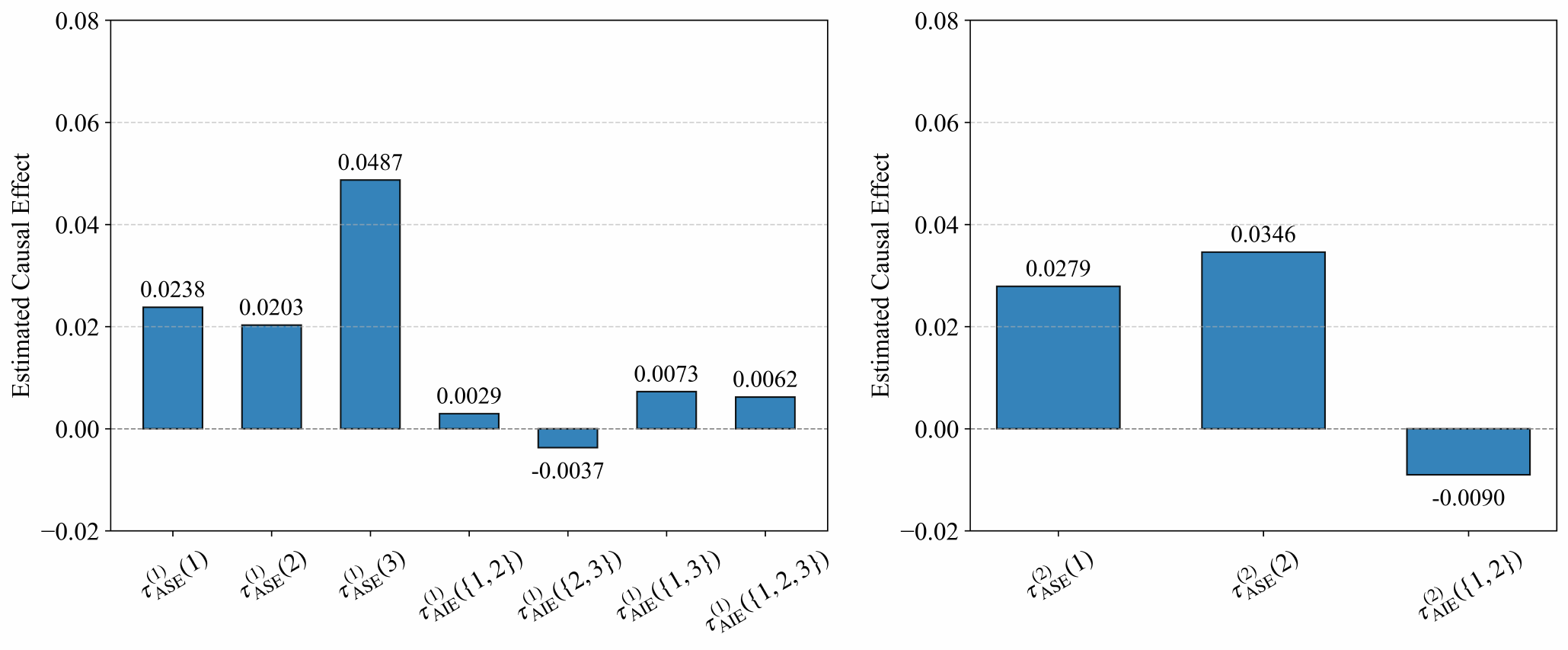}
    \caption{$\alpha=1.0$}
    \label{fig:empirical_results_alpha_1}
  \end{subfigure}

  \caption{Estimated single and interaction treatment effects obtained from two real-world marketing promotion datasets under different balancing-penalty weights.
The top panel corresponds to $\alpha=0.5$, and the bottom panel corresponds to $\alpha=1.0$.
In each panel, the left side shows results for dataset~1~(CP\textsubscript{1}--CP\textsubscript{3}), and the right side shows results for dataset~2~(CP\textsubscript{4} and CP\textsubscript{5}).
Here, $\smash{\tau_{\mathrm{ASE}}^{(d)}(k)}$ and $\smash{\tau_{\mathrm{AIE}}^{(d)}(S)}$ denote the estimated effects for single and multiple treatments in dataset $d$.
The outcome is standardized prior to estimation.}
  \label{fig:empirical_results_alpha_sensitivity}
\end{figure}

This appendix reports how the estimated causal effects change when varying the balancing-penalty weight $\alpha$ in the two real-world datasets derived from multiple marketing promotions. 
In real-world observational data, the data-generating process is unknown, and therefore, it is unclear whether the hyperparameter configuration that performs best in simulation settings is optimal for empirical analyses.
Among the various hyperparameters in CISI-Net that control different components of the model, we focus on the balancing-penalty weight $\alpha$, which is a core component of the proposed method.
By changing $\alpha$, we evaluate whether the qualitative conclusions of causal effect estimation depend excessively on a specific hyperparameter setting.

Figure~\ref{fig:empirical_results_alpha_sensitivity} shows the estimated causal effects obtained when the balancing-penalty weight is set to $\alpha=0.5$ and $\alpha=1.0$. 
A comparison with the results reported in Figure~\ref{fig:empirical_result} demonstrates that the magnitude of the estimated effects varies across these values.
At the same time, the sign of all estimated effects remains consistent and does not reverse. 
This result suggests that the qualitative conclusions of the real-world analysis are stable across variations in the balancing-penalty weight.

\end{document}